\documentclass[lettersize,journal]{IEEEtran}
\usepackage{amsmath,amsfonts}
\usepackage{algorithmic}
\usepackage{algorithm}
\usepackage{array}
\usepackage[caption=false,font=normalsize,labelfont=sf,textfont=sf]{subfig}
\usepackage{textcomp}
\usepackage{stfloats}
\usepackage{url}
\usepackage{verbatim}
\usepackage{graphicx}
\usepackage{cite}
\usepackage{caption}
\usepackage{color,soul}
\usepackage{float}
\usepackage{tikz}
\usepackage{pgfplots}
\usepackage{pgfplotstable}
\pgfplotsset{compat=1.18}
\usepackage{pgf-pie}
\usetikzlibrary{patterns}
\usepackage[dvipsnames]{xcolor}
\usetikzlibrary{plotmarks}
\usepackage{multirow}
\usepackage{makecell}
\usepackage{booktabs}

\usepackage{etoolbox}
\usepackage[colorlinks=true, linkcolor=blue, citecolor=blue, urlcolor=blue]{hyperref}

\makeatletter
\renewcommand{\@cite}[2]{%
  \leavevmode
  \begingroup
  \def\@temp##1{%
    \fcolorbox{green}{white}{##1}%
  }%
  [\@for\@cite@temp:=#1\do{%
    \@temp{\@cite@temp}%
  }]%
  \endgroup
}
\makeatother
\makeatletter
\let\old@ref\ref
\newcommand{\figboxref}[1]{\fcolorbox{cyan}{white}{Figure~\ref{#1}}}
\newcommand{\tabboxref}[1]{\fcolorbox{orange}{white}{Table~\ref{#1}}}

\makeatother

\begin{document}

\title{Revision-Aware Success Prediction from Multi-Attempt Programming Trajectories}

\author{Md Faizul Ibne Amin,~\IEEEmembership{Member,~IEEE,} Yutaka Watanobe,~\IEEEmembership{Member,~IEEE,} Daniel M. Muepu,~\IEEEmembership{Graduate Student Member,~IEEE,} Kenta Nanaumi, Haruto Suzuki, Md. Shahajada Mia, and Md Mostafizer Rahman}

\maketitle

\begin{abstract}
Programming outcome prediction plays a central role in data-driven programming education, supporting learner modeling, timely intervention, and adaptive assistance. Yet predicting submission success is difficult due to heterogeneous error states, short-term revisions, and uneven future-horizon availability in programming trajectories. This study examines three prediction tasks under a unified formulation: whether the current attempt is accepted (Task~1), whether the next attempt is accepted (Task~2), and whether acceptance is reached within a three-attempt recovery window (Task~3). Each task is evaluated across current-only, pairwise, and multi-step input regimes using machine learning (ML), deep learning (DL), and transformer-based pretrained code models (PTMs), represented by \textit{LinearSVM}, \textit{XGBoost}, \textit{BiGRU}, \textit{BiLSTM}, \textit{GraphCodeBERT}, and \textit{CodeT5+}. Results show a consistent pattern: the current-only regime is the most reliable, while pairwise and multi-step history provide no consistent gain. ML models are the strongest and most stable overall, particularly in Tasks~1 and~3, and Task~2 is the hardest across all model families. DL and PTMs perform well on Task~3 but are more task-dependent. In the Task~3 current-only setting, \textit{XGBoost} achieves AP/PR-AUC of 99.09\% and MCC of 0.6325, while \textit{GraphCodeBERT} and \textit{CodeT5+} reach $F_1$ scores of 80.00\% and 73.68\%, respectively. A sensitivity analysis confirms that Task~3 conclusions hold most robustly for ML models under stricter future-horizon control. Across all settings, feature-based ML models remain highly effective for programming success prediction, while complex models offer value in specific settings. This work provides a systematic comparison across predictive formulations and offers robust modeling guidance for submission-aware analytics in programming education, where near-future success prediction can inform timely intervention in online judge (OJ) platforms and adaptive programming support systems.
\end{abstract}


\section{Introduction}
\IEEEPARstart{P}{rogramming} is a central activity in computer science education, software engineering (SE) practice, and automated assessment environments~\cite{amin2024multi, chen2020analysis}. In programming courses, OJ, and contest platforms, learners repeatedly submit code, receive feedback, revise their solutions, and gradually move toward accepted implementations~\cite{watanobe2022online, kiesmuller2009diagnosing}. These submission traces provide a rich source of process-level evidence about how code evolves, how errors are understood and corrected, and how success emerges over time~\cite{amin2026error, qian2024enhanced, chen2024detecting, amin2026llm}. Consequently, programming success prediction has become an important problem in educational data mining and learning analytics, as it can support learner modeling, timely intervention, adaptive assistance, and the design of more responsive programming support systems~\cite{mao2019one, li2025deep, cui2022tri}.

Despite this potential, predicting programming success from submission trajectories remains challenging. A programming attempt may fail because of syntax errors, wrong algorithmic choices, incomplete edge-case handling, or minor implementation mistakes~\cite{bouallegue2026machine, amin2025source}. A learner may produce several revisions before reaching acceptance, and these revisions may carry useful signals about debugging direction, implementation stability, or proximity to success~\cite{carter2015normalized, pires2024predicting}. Prior studies have shown that programming behavior, compilation histories, and learning trajectories can support prediction of student performance or future correctness~\cite{jiang2020knowledge, yang2025difficulty, shi2022code}. This line of work is closely related to programming knowledge tracing (KT), where sequential interactions estimate learner progress and predict future correctness~\cite{piech2015deep, wang2023dynamic}. However, most KT studies emphasize latent skill estimation or exercise-level future correctness~\cite{liu2019ekt, shen2024survey}, while this study focuses on attempt-level success prediction from code representations. The comparative value of short revision history in particular remains insufficiently understood when evaluated directly against the current code state under matched task definitions and evaluation conditions.

A key methodological question is whether revision-aware representations actually improve attempt-level success prediction, and which model family handles trajectory-based educational data most reliably. Adding revision history increases input length, reduces eligible samples, and may introduce noise; yet a current-code model, while simpler and more stable, may miss signals about debugging direction. Feature-based ML models can be competitive under limited data~\cite{chen2016xgboost}, while DL and PTMs such as \textit{GraphCodeBERT} and \textit{CodeT5+} offer stronger code representations but do not automatically outperform simpler models on small educational datasets~\cite{hochreiter1997long, guo2020graphcodebert, wang2023codet5+}. These questions are examined using judge-system submission trajectories from an AI-permitted human-AI co-creation contest (PCK finals)~\cite{PCK}, where participants revised code under competitive time constraints while using AI tools, preserving the iterative cycle of attempt, feedback, revision, and acceptance~\cite{amin2026llm}. Although the dataset is small, it reflects the realistic scale of contest-style programming events and provides a controlled setting for evaluating trajectory-based prediction under genuine data constraints. Three prediction horizons are studied: current-attempt success (Task~1), next-attempt success (Task~2), and near-future within a three-attempt recovery window (Task~3), each paired with current-only, pairwise, and multi-step input regimes. Each task corresponds to a distinct monitoring point: Task~1 supports real-time code assessment; Task~2 signals next-attempt intervention; and Task~3 identifies learners at risk of not recovering within a short revision window, potentially supporting submission-aware analytics in OJ platforms and adaptive programming systems.

The study is guided by the following research questions:
\begin{itemize}
    \item \textbf{RQ1:} To what extent can programming success be predicted from multi-attempt programming trajectories?
    \item \textbf{RQ2:} To what extent do revision-aware input regimes improve success prediction compared with current-only code representations?
    \item \textbf{RQ3:} How do ML, DL, and PTMs differ in predictive performance under a unified experimental protocol?
    \item \textbf{RQ4:} To what extent are the main findings robust to stricter trajectory-observation and data-construction constraints? 
\end{itemize}

The study compares \textit{LinearSVM} and \textit{XGBoost} (ML), \textit{BiGRU} and \textit{BiLSTM} (DL), and \textit{GraphCodeBERT} and \textit{CodeT5+} (PTM) under the same task-regime construction, trajectory-level split, validation-based selection, and test-based reporting protocol using complementary metrics covering ranking quality, thresholded decision, class-specific behavior, and probabilistic reliability. The results consistently favor current-only modeling and feature-based ML as the most reliable choices, with DL and PTMs providing complementary value in task-specific settings.

The contributions of this study are summarized as follows:
\begin{itemize}
    \item Judge-system submission trajectories from an AI-permitted human-AI co-creation contest (PCK finals) are utilized, where participants revised code under competitive time pressure while using AI tools. This setting represents an ecologically valid and rare real-world benchmark that simultaneously preserves automated verdict feedback, competitive time constraints, and AI tool use, which is not commonly found in existing programming prediction studies.

    \item Three attempt-level prediction horizons are introduced within a unified formulation, each grounded in a distinct potential OJ monitoring point, suggesting how prediction output could support actionable intervention in submission-aware programming analytics systems.

    \item Current-only, pairwise, and multi-step code representations are systematically compared under matched task definitions, split protocols, and evaluation conditions, providing direct evidence on whether and when revision history improves attempt-level success prediction.

    \item  A cross-family empirical evaluation across ML, DL, and PTMs under a unified protocol, together with a future-horizon sensitivity analysis, provides systematic performance evidence and modeling guidance for educational programming analytics under realistic data constraints.
\end{itemize}

The remainder of this paper is organized as follows. Section~\ref{sec:related_work} reviews related studies. Section~\ref{sec:motivation_task_design} introduces the motivation and task design. Section~\ref{sec:methodology} describes the data processing and modeling pipeline. Section~\ref{sec:experimental_setup} reports the experimental setup. Section~\ref{sec:experimental_results} presents the results, followed by the discussion and limitations in Sections~\ref{sec:discussion} and~\ref{sec:threats_validity}. Section~\ref{sec:conclusion} concludes the paper.

\section{Related Work}
\label{sec:related_work}
\subsection{Programming Success and Performance Prediction}
\label{subsec:rw_success_prediction}
Programming success and performance prediction have been studied from several educational data perspectives. Carter et al.~\cite{carter2015normalized} proposed the normalized programming state model, representing behavior as state transitions for computing-course performance prediction. Vives et al.~\cite{vives2024prediction} predicted pass/fail outcomes in a programming fundamentals course using LSTM and ML baselines, with SMOTE and GAN addressing class imbalance. Asthana et al.~\cite{asthana2023prediction} introduced trajectory-based learning coefficients and compared regression ML models for student performance prediction. Mao et al.~\cite{mao2019one} examined early success prediction in novice programming tasks using short observation windows, demonstrating the value of early behavioral signals. Pires et al.~\cite{pires2024predicting} analyzed trajectories for novice performance prediction, while Chen et al.~\cite{chen2024detecting} and Qian et al.~\cite{qian2024enhanced} studied automated analysis of programming behavior for educational support. Related work has also examined early risk detection, multi-source features, and course-level performance modeling~\cite{zhan2025predicting, wang2024multi, alhazmi2023early, alkan2025using}.

\subsection{Knowledge Tracing and Trajectory Modeling}
\label{subsec:rw_knowledge_tracing}
Piech et al.~\cite{piech2015deep} introduced deep KT using recurrent networks to model evolving student knowledge states. Shen et al.~\cite{shen2022monitoring} proposed LPKT and LPKT-S, modeling learning gain, forgetting, answer time, and student-specific progress to better align KT with the learning process. Zanellati et al.~\cite{zanellati2024hybrid} systematically reviewed 53 hybrid KT models by knowledge source, representation, and integration strategy. Zhu et al.~\cite{zhu2022programming} introduced a programming KT dataset and model for capturing programming learning behavior, and Shi et al.~\cite{shi2022code} incorporated source code into KT for student performance prediction. Yang et al.~\cite{yang2025difficulty} explored difficulty-aware programming KT with LLMs, and Jiang et al.~\cite{jiang2020knowledge} analyzed programming paths to track student progress across tasks. Further studies cover deep, hyperbolic, open-ended, and trajectory-oriented KT~\cite{liu2025deep, li2025hyperbolic, bajwa2019analyzing, liu2022open}.

\subsection{Code Representation Models for Programming Activity}
\label{subsec:rw_code_representation}
Source code carries lexical patterns, sequential structure, and semantic cues that different model families exploit in distinct ways. Feature-based ML models provide strong lexical baselines: \textit{LinearSVM} applies margin-based classification over high-dimensional sparse features~\cite{cortes1995support}, while \textit{XGBoost} captures nonlinear interactions through gradient-boosted trees~\cite{chen2016xgboost}. Sequential DL models add temporal structure: GRU~\cite{cho2014learning} and LSTM~\cite{hochreiter1997long} model token dependencies, and their bidirectional variants \textit{BiGRU} and \textit{BiLSTM} incorporate context from both directions~\cite{schuster1997bidirectional}. PTMs further enrich code representation: CodeBERT uses bimodal pretraining over code and natural language~\cite{feng2020codebert}, GraphCodeBERT adds data-flow information~\cite{guo2020graphcodebert}, CodeT5 introduces identifier-aware encoder-decoder pretraining~\cite{wang2021codet5}, and CodeT5+ extends this with stronger code-oriented pretraining and scaling~\cite{wang2023codet5+}.

\subsection{Human-AI Programming Contexts}
\label{subsec:rw_human_ai_programming}
AI-assisted programming has become increasingly relevant in education and SE, with studies covering code generation, explanation, debugging, refinement, assessment, and error analysis~\cite{crupi2025effectiveness, wu2025humanevalcomm, amin2026error}. Recent work has moved toward interactive workflows where users inspect and revise AI-supported outputs~\cite{amin2026llm, honarvar2025question, hwang202580}, connecting to human-AI co-creation settings where programming is shaped by collaboration, feedback, and human oversight rather than full automation~\cite{rezwana2023designing, kadenhe2025human}. In this study, the AI-permitted co-creation setting provides the data context, while the analysis focuses on judge-system submission trajectories.

\subsection{Research Gap and Position of This Study}
\label{subsec:rw_position}
The reviewed studies confirm that programming behavior, submission records, learning trajectories, KT methods, and code representations are valuable for educational prediction. However, most prior work targets course-level performance, pass/fail outcomes, or latent knowledge states, whereas this study focuses on attempt-level success prediction from judge-system submissions. Trajectory and KT studies demonstrate the value of sequential evidence but do not directly compare the current code state against short revision histories under matched task definitions and split protocols. ML, DL, and PTMs have each been applied to programming prediction but are not always evaluated side by side under matched task-regime conditions. Prior trajectory studies also rely predominantly on standard course or OJ datasets, without the AI-permitted contest setting examined here, where participants revise code under competitive constraints while using AI tools. Finally, near-future recovery prediction has not been examined with explicit future-horizon control, leaving the sensitivity of such tasks to data-construction choices unaddressed. This study addresses these gaps by defining three attempt-level success-prediction tasks over a rare real-world trajectory dataset, comparing current-only, pairwise, and multi-step representations across ML, DL, and PTM families, and introducing a sensitivity analysis for stricter future-horizon availability.

\section{Motivation and Task Design}
\label{sec:motivation_task_design}
Programming education systems generate rich submission traces, yet most feedback a learner receives is binary: accepted or not. The question of whether these traces can support more informative, attempt-level monitoring motivates the task design in this study. Each attempt in a multi-attempt trajectory serves as a state from which prediction labels and input representations are constructed. Preprocessing, feature extraction, and model-specific implementation are described in Section~\ref{sec:methodology}; dataset statistics are reported in Section~\ref{sec:experimental_setup}.

\subsection{Motivation for Revision-Aware Success Prediction}
\label{subsec:motivation_revision_aware}
\begin{figure}[t]
\centering
\includegraphics[width=3.65in]{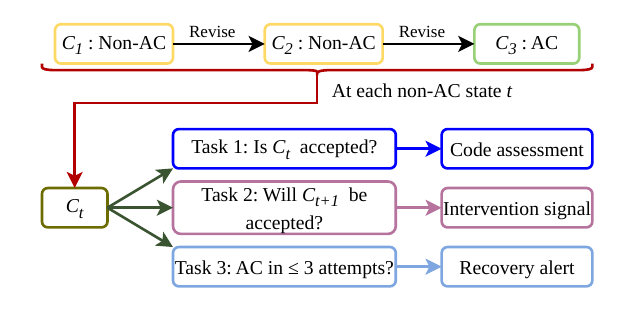}
\caption{Three prediction tasks over a programming trajectory and their potential monitoring outputs. Verdicts are ground-truth labels shown for context; at inference, the model predicts them from the submitted code state.}
\label{fig:motivational_example}
\end{figure}

Consider a learner solving a programming problem on an OJ. After each failed submission, they revise their code and resubmit, forming a trajectory of attempts in which prior code states may carry signals about debugging direction and proximity to success.\figboxref{fig:motivational_example} illustrates this iterative cycle and shows how the three prediction tasks correspond to distinct monitoring points within it. Task~1 assesses whether the current code is likely to be accepted, providing a real-time signal about code correctness at the moment of submission. Task~2 predicts whether the learner's next revision will succeed, indicating whether the current debugging direction appears productive. Task~3 takes a broader view: it estimates whether acceptance will occur within a three-attempt window, which is particularly useful for identifying learners who may remain stuck and could benefit from timely support, such as a hint, a targeted suggestion, or an instructor alert. This motivates revision-aware representations that incorporate one or more preceding attempts alongside the current code. At the same time, revision history is not guaranteed to help: it increases input length, may introduce noise, and reduces eligible samples since early trajectory states lack prior context. A simpler current-only model may be more practical for real-time deployment while still providing reliable predictions. This study therefore compares current-only, pairwise, and multi-step representations directly rather than assuming revision-aware inputs are superior.

\subsection{Multi-Attempt Programming Trajectories}
\label{subsec:programming_trajectories}
A programming trajectory is defined as an ordered sequence of submission attempts by a user for a single problem:
\begin{equation}
\mathcal{T}_{u,p} =
\left[
(C_1, v_1),
(C_2, v_2),
\ldots,
(C_n, v_n)
\right],
\end{equation}

where \(u\) is the user, \(p\) is the problem, \(C_t\) is the submitted code 
at attempt \(t\), \(v_t\) is the corresponding verdict, and \(n\) is the total 
number of attempts. The accepted verdict is denoted \(AC\); all other verdicts 
are treated as unsuccessful for binary prediction. Each attempt 
\(t \in \{1,\ldots,n\}\) defines a trajectory state from which task labels and 
input representations are constructed.

\subsection{Prediction Task Design}
\label{subsec:prediction_task_design}
Three binary prediction tasks capture different prediction horizons and levels of difficulty.

\subsubsection{Task 1: Current-Attempt Acceptance}
\label{subsubsec:task1}
For attempt \(t\), the label is defined as:
\begin{equation}
y_t^{(1)} = \mathbf{1}[v_t = AC],
\end{equation}

where \(\mathbf{1}[\cdot]\) is the indicator function. This is the most direct setting, as the label corresponds to the current code state.

\subsubsection{Task 2: Next-Attempt Acceptance}
\label{subsubsec:task2}
Task~2 predicts whether the next submission will be accepted:
\begin{equation}
y_t^{(2)} = \mathbf{1}[v_{t+1} = AC].
\end{equation}

The target depends on the learner's next revision rather than the current verdict, making this task harder than Task~1. States without a subsequent attempt are excluded since \(v_{t+1}\) is unobserved.

\subsubsection{Task 3: Near-Future Acceptance}
\label{subsubsec:task3}
Let \(h_t = n - t\) be the number of future attempts available after state 
\(t\). The label is:
\begin{equation}
y_t^{(3)} =
\mathbf{1}
\left[
\exists j \in \{1,2,3\},\ j \leq h_t
\ \text{such that}\
v_{t+j} = AC
\right].
\end{equation}

This captures near-future recovery rather than immediate success. States near the end of a trajectory may have fewer than three observable future attempts, creating unequal future-horizon availability across samples.

\subsection{Input Regimes and Data Representation}
\label{subsec:input_regimes}
Three input regimes determine which code states are provided to the model.

\subsubsection{Current-Only Representation}
\label{subsubsec:current_only}
The current-only regime uses only the current submission:
\begin{equation}
x_t^{\text{cur}} = C_t.
\end{equation}

This tests whether the current code snapshot alone is sufficient for success prediction.

\subsubsection{Pairwise Revision Representation}
\label{subsubsec:pairwise}
The pairwise regime uses the previous and current code states:
\begin{equation}
x_t^{\text{pair}} = (C_{t-1}, C_t).
\end{equation}

It is available only when at least one prior attempt exists and captures the immediate revision transition. Model-specific encoding with state markers is described in Section~\ref{sec:methodology}.

\subsubsection{Multi-Step Revision Representation}
\label{subsubsec:multistep}
The multi-step regime uses the previous two and the current code state:
\begin{equation}
x_t^{\text{multi}} = (C_{t-2}, C_{t-1}, C_t).
\end{equation}

It requires at least two prior attempts and tests whether a short local history improves prediction, but also reduces eligible samples and increases input complexity.

\subsection{Task-Regime Construction}
\label{subsec:task_regime_construction}
Each task-regime combination is constructed from states satisfying the required 
current, prior, and future code-verdict availability. Task~1 requires \(v_t\); Task~2 requires \(v_{t+1}\); Task~3 uses available future attempts up to three steps. Pairwise and multi-step regimes additionally require \(C_{t-1}\) and 
\((C_{t-2}, C_{t-1})\), respectively. These constraints produce different sample sizes across combinations: current-only settings retain the most eligible states, while pairwise and multi-step settings are progressively smaller. This reflects the core trade-off: revision-aware inputs may carry richer temporal signals but reduce data availability.

\subsection{Sensitivity Design for Task 3}
\label{subsec:task3_sensitivity_design}
To examine whether Task~3 findings are sensitive to unequal future-horizon 
availability, a robustness analysis is defined using two stricter subsets based on \(h_t\):
\begin{equation}
\mathcal{D}_{\text{S1}} =
\{t : h_t \geq 1\},
\end{equation}
\begin{equation}
\mathcal{D}_{\text{S3}} =
\{t : h_t \geq 3\}.
\end{equation}

S1 retains states with at least one observable future attempt; S3 retains only states with the full three-step horizon. The Task~3 label definition is unchanged; only the eligible state set is restricted. If a stricter subset becomes single-class after filtering, it is treated as a data-availability observation rather than a predictive benchmark, since binary discrimination requires both positive and negative examples.

\section{Methodology}
\label{sec:methodology}

\subsection{Pipeline Overview}
\label{subsec:pipeline_overview}
\begin{figure*}[t]
\centering
\includegraphics[width=5.5in]{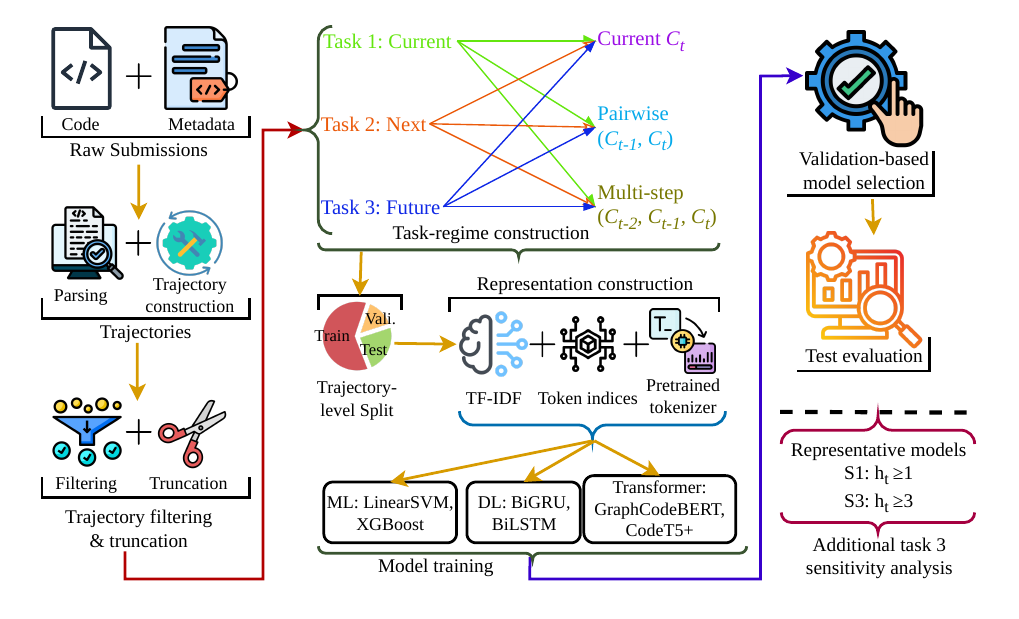}
\caption{Overview of the revision-aware success prediction workflow.}
\label{fig:approach_workflow}
\end{figure*}

\figboxref{fig:approach_workflow} summarizes the overall pipeline. Raw submission records are parsed into a canonical row-level table, reconstructed into participant-problem trajectories, filtered for revision validity, converted into task-regime datasets, and evaluated using ML, DL, and PTMs. For each task-regime-model setting, the validation split is used for checkpoint selection and threshold tuning; the test split is reserved for final performance reporting.

\subsection{Submission Parsing and Trajectory Reconstruction}
\label{subsec:trajectory_construction}
Each submission record is parsed into an attempt-level state by extracting the source code, verdict, timestamp, participant identifier, problem identifier, submission identifier, and programming language. Participant identifiers are obtained from the directory structure; submission and problem identifiers are parsed from the file naming convention. Source code is extracted from code containers, and verdicts and timestamps are identified from the visible record text. Two parsing steps are critical for valid trajectory reconstruction. First, verdicts are normalized using structured result fields before fallback keyword matching; both ``Accepted'' and ``Accept'' are mapped to \(AC\), while all other recognized verdicts are mapped to non-\(AC\). Second, timestamp extraction prioritizes the 
submission event timestamp, since page-level timestamps can corrupt chronological ordering and invalidate revision-based histories. After parsing, submissions are grouped by participant and problem, sorted chronologically, and assigned attempt indices. The reconstructed trajectory retains language metadata for filtering:
\begin{equation}
\mathcal{T}_{u,p}^{\ell}
=
\left[
(C_1, v_1, \ell_1),
(C_2, v_2, \ell_2),
\ldots,
(C_n, v_n, \ell_n)
\right],
\end{equation}

where \(C_t\) is the submitted source code, \(v_t\) is the normalized verdict, and \(\ell_t\) is the detected language.

\subsection{Language-Consistent Filtering and Trajectory Truncation}
\label{subsec:language_filtering_truncation}
Revision-aware regimes assume consecutive attempts are revisions of the same 
code artifact; when the language changes within a trajectory, transitions such 
as \(C_{t-1} \rightarrow C_t\) would mix revision effects with language-change 
effects. A trajectory is therefore retained only if it is language-consistent:
\begin{equation}
\ell_1 = \ell_2 = \cdots = \ell_n.
\end{equation}

This check is applied per participant-problem trajectory, so the dataset can contain multiple languages across trajectories while each retained trajectory remains monolingual. Unsupported languages with negligible frequency are also removed.

Successful trajectories are truncated at the first accepted submission. Let
\begin{equation}
t_{AC} = \min \{t : v_t = AC\}
\end{equation}
be the first accepted attempt if it exists. The retained trajectory is:
\begin{equation}
\widetilde{\mathcal{T}}_{u,p}
=
\begin{cases}
\left[(C_1,v_1,\ell_1),\ldots,(C_{t_{AC}},v_{t_{AC}},\ell_{t_{AC}})\right],
& \text{if } t_{AC} \text{ exists},\\[2mm]
\mathcal{T}_{u,p}^{\ell},
& \text{otherwise}.
\end{cases}
\end{equation}

This truncation prevents post-success attempts from influencing recovery-oriented prediction tasks; if no accepted attempt exists, the full trajectory is retained. For each state \(t\), the available future horizon is:
\begin{equation}
h_t = n_t - t,
\end{equation}

where \(n_t\) is the length of the retained trajectory containing state \(t\), which may differ from the original trajectory length \(n\) before truncation. This value is used for future-window construction and the Task~3 sensitivity analysis.

\subsection{Task-Specific Instance Construction}
\label{subsec:task_specific_instance_construction}
From each retained trajectory, supervised instances are constructed for every task-regime combination. For task \(k\) and regime \(r\):
\begin{equation}
\mathcal{D}_{k,r}
=
\left\{
(x_t^{(r)}, y_t^{(k)}, m_t)
\right\}_{t \in \Omega_{k,r}},
\end{equation}

where \(x_t^{(r)}\) is the regime-specific input, \(y_t^{(k)}\) is the task label, \(m_t\) contains metadata (trajectory identifier, participant, problem, language, attempt index), and \(\Omega_{k,r}\) is the eligible state set.

Task~2 and Task~3 are constructed from non-\(AC\) states, where future success prediction is meaningful. Current-only, pairwise, and multi-step regimes progressively reduce \(\Omega_{k,r}\) by requiring additional historical context. A trajectory-level train/validation/test split prevents data leakage: all states from the same participant-problem trajectory are assigned to the same split. The split is selected so that every main task-regime-split combination contains both positive and negative examples, enabling threshold tuning and valid binary evaluation. The Task~3 sensitivity datasets are constructed after the main task-regime datasets. S1 retains states with \(h_t \geq 1\); S3 retains states with \(h_t \geq 3\). Single-class subsets after filtering are reported as data-availability observations rather than predictive benchmarks.

\subsection{Input Representation Construction}
\label{subsec:representation_construction}
For each eligible state, inputs are constructed per regime using state markers 
\(M_0\), \(M_1\), \(M_2\) to separate code snapshots:
\begin{equation}
x_t^{(\mathrm{cur})} = C_t,
\end{equation}
\begin{equation}
x_t^{(\mathrm{pair})}
=
[M_1; C_{t-1}; M_0; C_t],
\end{equation}
\begin{equation}
x_t^{(\mathrm{multi})}
=
[M_2; C_{t-2}; M_1; C_{t-1}; M_0; C_t].
\end{equation}

The markers preserve the temporal role of each code state. Each model family converts \(x_t^{(r)}\) into family-specific features: ML models use sparse lexical features, DL models use token-index sequences from a train-only vocabulary, and PTMs apply the model's tokenizer with truncation and padding. Learned components such as TF-IDF vectorizers and DL vocabularies are fitted on the training split only.

\subsection{Machine Learning Pipeline}
\label{subsec:ml_pipeline}
The ML pipeline uses TF-IDF features over the constructed representation. For each task-regime dataset, a vectorizer \(\phi_r(\cdot)\) is fitted on the training split:
\begin{equation}
\mathbf{z}_t^{(r)} = \phi_r(x_t^{(r)}).
\end{equation}

\textit{LinearSVM} provides a sparse linear baseline; \textit{XGBoost} provides 
a nonlinear tree-based alternative. Each model maps the feature vector to a 
positive-class score:
\begin{equation}
s_t = f_{\theta}(\mathbf{z}_t^{(r)}),
\end{equation}

where \(s_t\) is used for ranking-based analysis and the binary prediction uses a validation-selected threshold \(\tau\):
\begin{equation}
\hat{y}_t = \mathbf{1}[s_t \geq \tau].
\end{equation}

\subsection{Deep Learning Pipeline}
\label{subsec:dl_pipeline}
The DL pipeline represents each input as a token-index sequence. A vocabulary \(V\) is built from the training split only. For input \(x_t^{(r)}\), tokenization produces:
\begin{equation}
\mathbf{q}_t = [q_1, q_2, \ldots, q_L],
\end{equation}

where \(L\) is the maximum sequence length after truncation or padding. Token  indices are mapped to learned embeddings:
\begin{equation}
\mathbf{E}_t = [\mathbf{e}_1, \mathbf{e}_2, \ldots, \mathbf{e}_L].
\end{equation}

Two bidirectional recurrent architectures are evaluated: \textit{BiGRU} and 
\textit{BiLSTM}. The encoder produces forward and backward summaries, concatenated as:
\begin{equation}
\mathbf{h}_t =
[\overrightarrow{\mathbf{h}}_t ; \overleftarrow{\mathbf{h}}_t].
\end{equation}

The binary logit and positive-class score are:
\begin{equation}
o_t = \mathbf{w}^{\top}\mathrm{Dropout}(\mathbf{h}_t) + b,
\end{equation}
\begin{equation}
s_t = \sigma(o_t).
\end{equation}

Class-weighted binary cross-entropy addresses imbalance. The positive-class 
weight is:
\begin{equation}
\alpha =
\frac{N_{\mathrm{neg}}^{\mathrm{train}}}
{N_{\mathrm{pos}}^{\mathrm{train}}},
\end{equation}

where \(N_{\mathrm{pos}}^{\mathrm{train}}\) and \(N_{\mathrm{neg}}^{\mathrm{train}}\) denote the number of positive and negative training instances. The per-instance loss is:
\begin{equation}
\mathcal{L}_t
=
-\alpha y_t \log(s_t)
-
(1-y_t)\log(1-s_t).
\end{equation}

\subsection{Transformer-Based Pretrained Code Model Pipeline}
\label{subsec:transformer_pipeline}
Two PTMs are evaluated: \textit{GraphCodeBERT} and \textit{CodeT5+}. The pretrained encoder produces contextual hidden states:
\begin{equation}
\mathbf{H}_t =
\mathrm{Enc}_{\theta}(x_t^{(r)})
=
[\mathbf{h}_{t,1}, \mathbf{h}_{t,2}, \ldots, \mathbf{h}_{t,L}].
\end{equation}

For \textit{GraphCodeBERT}, the first-token hidden state serves as the sequence-level representation:
\begin{equation}
\mathbf{c}_t = \mathbf{h}_{t,1}.
\end{equation}

For \textit{CodeT5+}, masked mean pooling is applied over encoder hidden states:
\begin{equation}
\mathbf{c}_t =
\frac{\sum_{i=1}^{L} a_i \mathbf{h}_{t,i}}
{\sum_{i=1}^{L} a_i},
\end{equation}

where \(a_i \in \{0,1\}\) is the attention-mask value. The logit and 
positive-class score are then computed from \(\mathbf{c}_t\) as:
\begin{equation}
o_t = \mathbf{w}^{\top}\mathrm{Dropout}(\mathbf{c}_t) + b,
\end{equation}
\begin{equation}
s_t = \sigma(o_t).
\end{equation}

State markers are included in pairwise and multi-step inputs so the model distinguishes previous and current code states.

\subsection{Training, Selection, and Evaluation Protocol}
\label{subsec:training_selection_protocol}
Training uses the training split only. For DL and PTM models, the best checkpoint is selected using validation AP/PR-AUC, which is threshold-independent and emphasizes positive-class ranking. ML models require no epoch-level checkpointing but use validation scores for threshold selection. The decision threshold is then chosen on the validation split by maximizing MCC:
\begin{equation}
\tau^{*}
=
\arg\max_{\tau \in \mathcal{G}}
\mathrm{MCC}
\left(
\mathbf{y}^{\mathrm{val}},
\mathbf{1}[\mathbf{s}^{\mathrm{val}} \geq \tau]
\right),
\end{equation}

where \(\mathcal{G}\) is the candidate threshold grid, \(\mathbf{y}^{\mathrm{val}}\) 
is the vector of validation labels, and \(\mathbf{s}^{\mathrm{val}}\) is the 
vector of validation scores. The selected \(\tau^{*}\) is fixed and applied 
to the test split:
\begin{equation}
\hat{y}_t^{\mathrm{test}}
=
\mathbf{1}[s_t^{\mathrm{test}} \geq \tau^{*}].
\end{equation}

Final performance is evaluated on the test split; the metric set and reporting structure are described in Section~\ref{sec:experimental_setup}.

\section{Experimental Setup}
\label{sec:experimental_setup}
\subsection{Data Source and Original Submission Corpus}
\label{subsec:data_source}
The dataset is derived from the AI-permitted co-creation track conducted in parallel with the official PCK finals~\cite{PCK}. In this setting, 15 participants attempted 13 contest problems under the same time constraints while being allowed to use AI tools. Although prompt logs were collected in the broader event, the present study focuses only on judge-system submission records, including source code, verdicts, timestamps, programming language, and participant-problem identifiers. The original corpus contains 517 submission attempts across 184 participant-problem trajectories \((u,p)\), reconstructed into chronological trajectories and filtered for language consistency.

\subsection{Task-Regime Dataset Summary}
\label{subsec:task_regime_dataset_summary}
After parsing, filtering, truncation, and task-regime eligibility construction, nine main task-regime datasets were generated, each split into training, validation, and test sets using a trajectory-level partition. \tabboxref{tab:dataset_task_regime_summary} summarizes the split-wise class 
distribution for the final modeling dataset.\footnote{Cur, Pair, and Mul denote current-only, pairwise, and multi-step regimes. Val is the validation.\(+\) and \(-\) denote positive and negative class counts.}

\begin{table}[h]
\centering
\caption{Split-wise class distribution for task-regime datasets}
\setlength{\arrayrulewidth}{0.3mm}
\setlength{\tabcolsep}{8.5pt}
\renewcommand{\arraystretch}{1}
\label{tab:dataset_task_regime_summary}
\scriptsize
\begin{tabular}{l||l||l||l||l}
\hline
\textbf{Task} & \textbf{Regime} & \textbf{Train $+/-$} & \textbf{Val $+/-$} & \textbf{Test $+/-$} \\ \hline \hline
\multirow{3}{*}{Task$\sim$1} & Cur & 209 (102/107) & 44 (21/23) & 40 (22/18) \\
 & Pair & 98 (22/76) & 20 (7/13) & 16 (6/10) \\
 & Mul & 68 (13/55) & 11 (2/9) & 9 (3/6) \\ \hline
\multirow{3}{*}{Task$\sim$2} & Cur & 98 (22/76) & 20 (7/13) & 16 (6/10) \\
 & Pair & 68 (13/55) & 11 (2/9) & 9 (3/6) \\
 & Mul & 49 (6/43) & 7 (1/6) & 5 (1/4) \\ \hline
\multirow{3}{*}{Task$\sim$3} & Cur & 107 (41/66) & 23 (10/13) & 18 (10/8) \\
 & Pair & 76 (24/52) & 13 (4/9) & 10 (4/6) \\
 & Mul & 55 (15/40) & 9 (2/7) & 6 (1/5) \\ \cline{2-3} \hline \hline
\end{tabular}
\end{table}

For the Task~3 sensitivity analysis, \tabboxref{tab:task3_sensitivity_dataset_summary} reports the class distribution of the stricter horizon subsets. S1 remains valid for binary evaluation; S3 becomes single-class after filtering and is treated as a data-availability observation.

\begin{table}[h]
\centering
\caption{Class distribution for Task~3 sensitivity subsets}
\setlength{\arrayrulewidth}{0.3mm}
\setlength{\tabcolsep}{11.7pt}
\renewcommand{\arraystretch}{1}
\label{tab:task3_sensitivity_dataset_summary}
\begin{tabular}{l||c||c||c}
\toprule
\textbf{Subset} & \textbf{Train $+/-$} & \textbf{Val $+/-$} & \textbf{Test $+/-$} \\ \hline
\midrule
S1: \(h_t \geq 1\) & 76 (19/57) & 13 (3/10) & 10 (4/6) \\
S3: \(h_t \geq 3\) & 43 (0/43)  & 6 (0/6)  & 4 (0/4) \\ \hline
\bottomrule
\end{tabular}
\end{table}

\subsection{Evaluation Metrics}
\label{subsec:evaluation_metrics}
The experiments use complementary metrics since the datasets are often class-imbalanced and the decision threshold is validation-selected. \tabboxref{tab:evaluation_metrics_summary} summarizes the metric groups.\footnote{AP/PR-AUC refers to average precision computed over the 
precision-recall curve.}

\begin{table}[h]
\centering
\caption{Evaluation metrics and their interpretation}
\setlength{\arrayrulewidth}{0.3mm}
\setlength{\tabcolsep}{2.5pt}
\renewcommand{\arraystretch}{1}
\label{tab:evaluation_metrics_summary}
\scriptsize
\begin{tabular}{l||l||l}
\toprule
\textbf{Metric} & \textbf{Role} & \textbf{Preferred direction} \\ \hline
\midrule
AP/PR-AUC & Precision--recall ranking quality & Higher is better \\ \hline
ROC-AUC & Class-separation ability & Higher is better \\ \hline
MCC & Balanced thresholded decision quality & Higher is better \\ \hline
$F_1$ & Harmonic mean of precision and recall & Higher is better \\ \hline
\shortstack{Balanced \\Accuracy ($BA$)} & Mean of recall and specificity & Higher is better \\ \hline
Precision ($P$) & Reliability of positive predictions & Higher is better \\ \hline
Recall ($R$) & Coverage of positive cases & Higher is better \\ \hline
Specificity ($TNR$) & Coverage of negative cases & Higher is better \\ \hline
Brier Score ($BS$) & Probabilistic error & Lower is better \\ \hline
Log Loss ($\mathcal{L}_{\log}$) & Probabilistic calibration penalty & Lower is better \\ \hline
Threshold ($\tau$) & Validation-selected decision cutoff & Reported value \\ \hline
\bottomrule
\end{tabular}
\end{table}

\subsection{Hyperparameter Settings}
\label{subsec:hyperparameter_settings}
Hyperparameters were kept fixed within each model family across all 
task-regime combinations to focus the comparison on task definition, input 
regime, and model family. \tabboxref{tab:hyperparameter_summary} summarizes 
the settings. Early stopping was applied to all neural models. DL models trained for up to 20 epochs with patience 5; PTMs used 10 epochs with patience 2, since pretrained encoders converge faster during fine-tuning. The best checkpoint was selected by validation AP/PR-AUC in both cases. For the Task~3 sensitivity analysis, one representative per family was chosen based on main Task~3 results: \textit{XGBoost} (ML), \textit{BiLSTM} (DL), and \textit{GraphCodeBERT} (PTM).

\begin{table*}[t]
\centering
\caption{Hyperparameter settings used in the experiments}
\setlength{\arrayrulewidth}{0.2mm}
\setlength{\tabcolsep}{11.8pt}
\renewcommand{\arraystretch}{1.1}
\label{tab:hyperparameter_summary}
\scriptsize
\begin{tabular}{l||l||l}
\hline
\textbf{Family} & \textbf{Component} & \textbf{Setting} \\ \hline \hline
\multirow{4}{*}{\shortstack{ML \\(\textit{LinearSVM}, \textit{XGBoost})}}
& Feature representation & TF-IDF over constructed code representation \\ \cline{2-3}
& TF-IDF features & Character word-boundary ; 3--5 grams ; \texttt{min\_df}=1, \texttt{max\_features}=50k, sublinear TF \\ \cline{2-3}
& \textit{LinearSVM} & \shortstack{\(C=1.0\), balanced class weight, max iterations = 10k ; sigmoid calibration (3-fold CV)} \\ \cline{2-3}
& \textit{XGBoost} & 300 trees, max depth = 4, learning rate = 0.05 ; subsample=colsample\_bytree=0.9, \(\lambda=1.0\) \\ \cline{2-3} \hline \hline

\multirow{6}{*}{\shortstack{DL \\(\textit{BiGRU}, \textit{BiLSTM})}}
& Tokenization & Regex-based code tokenizer \\ \cline{2-3}
& Vocabulary & Train-only vocabulary ; max size = 20k ; min frequency = 1 \\ \cline{2-3}
& Sequence length / batch size & 512 tokens ; 16 \\ \cline{2-3}
& Architecture & Embedding=128 ; hidden dim.=128 ; 1 layer; dropout=0.30 \\ \cline{2-3}
& Optimizer & AdamW; learning rate = \(1\mathrm{e}{-3}\) ; weight decay = \(1\mathrm{e}{-4}\) \\ \cline{2-3}
& Loss & Class-weighted BCE using train-only positive weight \\ \cline{2-3} \hline \hline

\multirow{5}{*}{\shortstack{PTM \\(\textit{GraphCodeBERT}, \textit{CodeT5+})}}
& Tokenization & Model-specific pretrained tokenizer \\ \cline{2-3}
& Maximum length / batch size & 256 tokens ; 8 \\ \cline{2-3}
& Optimizer & AdamW; learning rate = \(2\mathrm{e}{-5}\) ; Weight decay / warmup = \(1\mathrm{e}{-4}\) / 0.10  \\ \cline{2-3}
& Dropout / gradient clipping & 0.30 ; 1.0 \\ \cline{2-3}
& Loss / precision & Class-weighted BCE; mixed precision enabled when CUDA is available \\ \hline \hline
\end{tabular}
\end{table*}

\subsection{Device Configuration}
\label{subsec:device_configuration}
All neural experiments used a fixed random seed of 44 and ran on an NVIDIA 
GeForce RTX 2080 Ti via CUDA.

\section{Results}
\label{sec:experimental_results}
The results are reported task-wise to preserve the experimental structure. For each task, test performance is presented across three metric-grouped tables: ranking metrics (AP/PR-AUC and ROC-AUC), thresholded decision metrics (MCC, $F_1$, and $BA$), and class-specific and probabilistic metrics ($P$, $R$, $TNR$, $BS$, $\mathcal{L}_{\log}$, and $\tau$). Confusion counts are not tabulated but are discussed where needed to explain prediction collapse or strong $P$-$R$ imbalance.

\subsection{Task 1: Current-Attempt Success Prediction}
\label{subsec:task1_results}
Task~1 predicts current-attempt acceptance. Ranking, thresholded decision, and class-specific results are in \tabboxref{tab:task1_ranking}, \tabboxref{tab:task1_decision}, and \tabboxref{tab:task1_class_calibration}, respectively. 

\begin{table}[h]
\centering
\caption{Task~1: ranking-oriented performance}
\setlength{\arrayrulewidth}{0.2mm}
\setlength{\tabcolsep}{12.7pt}
\renewcommand{\arraystretch}{1.1}
\label{tab:task1_ranking}
\scriptsize
\begin{tabular}{l||l||c||c}
\hline
\multicolumn{2}{c||}{\textbf{Model Configuration}} & \multicolumn{2}{c}{\textbf{Task 1}} \\ \hline
\multicolumn{1}{l|}{\textbf{Regime}} & \textbf{Model} & \multicolumn{1}{c|}{\textbf{AP/PR-AUC}} & \multicolumn{1}{c}{\textbf{ROC-AUC}} \\ \hline \hline
\multicolumn{1}{l|}{\multirow{6}{*}{Cur}} & \textit{LinearSVM} & \multicolumn{1}{l|}{0.8360} & 0.8131 \\ \cline{2-4} 
\multicolumn{1}{l|}{} & \textit{XGBoost} & \multicolumn{1}{l|}{\textbf{0.8775}} & 0.8586 \\ \cline{2-4} 
\multicolumn{1}{l|}{} & \textit{BiGRU} & \multicolumn{1}{l|}{0.5886} & 0.5328 \\ \cline{2-4} 
\multicolumn{1}{l|}{} & \textit{BiLSTM} & \multicolumn{1}{l|}{0.5891} & 0.5530 \\ \cline{2-4} 
\multicolumn{1}{l|}{} & \textit{GraphCodeBERT} & \multicolumn{1}{l|}{0.7915} & 0.7096 \\ \cline{2-4} 
\multicolumn{1}{l|}{} & \textit{CodeT5+} & \multicolumn{1}{l|}{0.8072} & 0.7197 \\ \hline \hline
\multicolumn{1}{l|}{\multirow{6}{*}{Pair}} & \textit{Linear SVM} & \multicolumn{1}{l|}{0.6605} & 0.7833 \\ \cline{2-4} 
\multicolumn{1}{l|}{} & \textit{XGBoost} & \multicolumn{1}{l|}{0.6218} & 0.8167 \\ \cline{2-4} 
\multicolumn{1}{l|}{} & \textit{BiGRU} & \multicolumn{1}{l|}{0.5200} & 0.5333 \\ \cline{2-4} 
\multicolumn{1}{l|}{} & \textit{BiLSTM} & \multicolumn{1}{l|}{0.4152} & 0.4833 \\ \cline{2-4} 
\multicolumn{1}{l|}{} & \textit{GraphCodeBERT} & \multicolumn{1}{l|}{0.6279} & 0.7167 \\ \cline{2-4} 
\multicolumn{1}{l|}{} & \textit{CodeT5+} & \multicolumn{1}{l|}{0.3580} & 0.4000 \\ \hline \hline
\multicolumn{1}{l|}{\multirow{6}{*}{Mul}} & \textit{LinearSVM} & \multicolumn{1}{l|}{0.6389} & 0.8333 \\ \cline{2-4} 
\multicolumn{1}{l|}{} & \textit{XGBoost} & \multicolumn{1}{l|}{0.8056} & \textbf{0.8889} \\ \cline{2-4} 
\multicolumn{1}{l|}{} & \textit{BiGRU} & \multicolumn{1}{l|}{0.3730} & 0.3333 \\ \cline{2-4} 
\multicolumn{1}{l|}{} & \textit{BiLSTM} & \multicolumn{1}{l|}{0.5778} & 0.5000 \\ \cline{2-4} 
\multicolumn{1}{l|}{} & \textit{GraphCodeBERT} & \multicolumn{1}{l|}{0.3472} & 0.3889 \\ \cline{2-4} 
\multicolumn{1}{l|}{} & \textit{CodeT5+} & \multicolumn{1}{l|}{0.5333} & 0.7222 \\ \hline \hline
\end{tabular}
\end{table}

Current-attempt acceptance is learnable from the submitted code state. The current-only regime provides the most stable ranking behavior: \textit{XGBoost} achieves the highest AP/PR-AUC (0.8775), while the highest ROC-AUC (0.8889) appears in the multi-step regime, again via \textit{XGBoost}. The multi-step result covers a smaller test subset, so it warrants more caution than the current-only figure.

\textit{XGBoost} gives the strongest ML pattern for Task~1. In the current-only regime, it reaches AP/PR-AUC of 0.8775, MCC of 0.5484, and $F_1$ of 0.7907. \textit{LinearSVM} behaves more conservatively, achieving the highest $P$ (0.8750) and strong $TNR$ (0.8889) in the same regime. In the pairwise regime, \textit{XGBoost} obtains the best thresholded result across Task~1: MCC of 0.7746, $F_1$ of 0.8571, and $BA$ of 0.9, suggesting pairwise revision information can help the nonlinear model at the selected threshold, though the best AP/PR-AUC still comes from current-only. In the multi-step regime, \textit{XGBoost} achieves the best $BS$ (0.1477) and $\mathcal{L}_{\log}$ (0.4199), but its very low selected threshold reflects an $R$-oriented operating point; the current-only result remains the more stable reference as it covers the largest subset.

The recurrent DL models produce weaker and less stable results than the ML models. Their best performance is in the current-only setting, where \textit{BiLSTM} reaches MCC of 0.2257 and $F_1$ of 0.5789. Neither pairwise nor multi-step regimes improve the recurrent models: both \textit{BiGRU} and \textit{BiLSTM} show weak thresholded performance in the pairwise regime, and the multi-step regime remains limited, though \textit{BiLSTM} outperforms \textit{BiGRU}. Under the present dataset size and from-scratch recurrent setup, revision history provides no reliable benefit for Task~1.

\begin{table}[h]
\centering
\caption{Task~1: thresholded decision performance}
\setlength{\arrayrulewidth}{0.2mm}
\setlength{\tabcolsep}{11.2pt}
\renewcommand{\arraystretch}{1.1}
\label{tab:task1_decision}
\scriptsize
\begin{tabular}{l||l||c||c||c}
\hline
\multicolumn{2}{c||}{\textbf{Model Configuration}} & \multicolumn{3}{c}{\textbf{Task 1}} \\ \hline
\multicolumn{1}{l|}{\textbf{Regime}} & \textbf{Model} & \multicolumn{1}{l|}{\textbf{MCC}} & \multicolumn{1}{l|}{\textbf{$F_1$}} & \textbf{$BA$} \\ \hline \hline
\multicolumn{1}{l|}{\multirow{6}{*}{Cur}} & \textit{LinearSVM} & \multicolumn{1}{l|}{0.5334} & \multicolumn{1}{l|}{0.7368} & 0.7626 \\ \cline{2-5} 
\multicolumn{1}{l|}{} & \textit{XGBoost} & \multicolumn{1}{l|}{0.5484} & \multicolumn{1}{l|}{0.7907} & 0.7753 \\ \cline{2-5} 
\multicolumn{1}{l|}{} & \textit{BiGRU} & \multicolumn{1}{l|}{0.0316} & \multicolumn{1}{l|}{0.4444} & 0.5152 \\ \cline{2-5} 
\multicolumn{1}{l|}{} & \textit{BiLSTM} & \multicolumn{1}{l|}{0.2257} & \multicolumn{1}{l|}{0.5789} & 0.6111 \\ \cline{2-5} 
\multicolumn{1}{l|}{} & \textit{GraphCodeBERT} & \multicolumn{1}{l|}{0.2423} & \multicolumn{1}{l|}{0.5556} & 0.6162 \\ \cline{2-5} 
\multicolumn{1}{l|}{} & \textit{CodeT5+} & \multicolumn{1}{l|}{0.2845} & \multicolumn{1}{l|}{0.7083} & 0.6364 \\ \hline \hline
\multicolumn{1}{l|}{\multirow{6}{*}{Pair}} & \textit{LinearSVM} & \multicolumn{1}{l|}{0.5222} & \multicolumn{1}{l|}{0.7059} & 0.7500 \\ \cline{2-5} 
\multicolumn{1}{l|}{} & \textit{XGBoost} & \multicolumn{1}{l|}{\textbf{0.7746}} & \multicolumn{1}{l|}{\textbf{0.8571}} & \textbf{0.9000} \\ \cline{2-5} 
\multicolumn{1}{l|}{} & \textit{BiGRU} & \multicolumn{1}{l|}{-0.0667} & \multicolumn{1}{l|}{0.3333} & 0.4667 \\ \cline{2-5} 
\multicolumn{1}{l|}{} & \textit{BiLSTM} & \multicolumn{1}{l|}{-0.0413} & \multicolumn{1}{l|}{0.2222} & 0.4833 \\ \cline{2-5} 
\multicolumn{1}{l|}{} & \textit{GraphCodeBERT} & \multicolumn{1}{l|}{0.4472} & \multicolumn{1}{l|}{0.6667} & 0.7000 \\ \cline{2-5} 
\multicolumn{1}{l|}{} & \textit{CodeT5+} & \multicolumn{1}{l|}{-0.2928} & \multicolumn{1}{l|}{0.0000} & 0.4000 \\ \hline \hline
\multicolumn{1}{l|}{\multirow{6}{*}{Mul}} & \textit{LinearSVM} & \multicolumn{1}{l|}{0.5000} & \multicolumn{1}{l|}{0.6667} & 0.7500 \\ \cline{2-5} 
\multicolumn{1}{l|}{} & \textit{XGBoost} & \multicolumn{1}{l|}{0.3780} & \multicolumn{1}{l|}{0.6000} & 0.6667 \\ \cline{2-5} 
\multicolumn{1}{l|}{} & \textit{BiGRU} & \multicolumn{1}{l|}{-0.1581} & \multicolumn{1}{l|}{0.2857} & 0.4167 \\ \cline{2-5} 
\multicolumn{1}{l|}{} & \textit{BiLSTM} & \multicolumn{1}{l|}{0.1890} & \multicolumn{1}{l|}{0.4000} & 0.5833 \\ \cline{2-5} 
\multicolumn{1}{l|}{} & \textit{GraphCodeBERT} & \multicolumn{1}{l|}{0.0000} & \multicolumn{1}{l|}{0.0000} & 0.5000 \\ \cline{2-5} 
\multicolumn{1}{l|}{} & \textit{CodeT5+} & \multicolumn{1}{l|}{0.3162} & \multicolumn{1}{l|}{0.5714} & 0.6667 \\ \hline \hline
\end{tabular}
\end{table}

The PTMs pick up useful signal in selected regimes but do not consistently outperform the stronger ML results. In the current-only regime, \textit{CodeT5+} obtains AP/PR-AUC of 0.8072 and $F_1$ of 0.7083; \textit{GraphCodeBERT} reaches AP/PR-AUC of 0.7915 with stronger $TNR$. In the pairwise regime, \textit{GraphCodeBERT} gives the strongest PTM thresholded result: MCC of 0.4472 and $F_1$ of 0.6667. \textit{CodeT5+} fails to identify positive cases in the pairwise setting, and \textit{GraphCodeBERT} collapses to an all-negative pattern in multi-step. The multi-step \textit{CodeT5+} result is more usable (MCC of 0.3162, $F_1$ of 0.5714) but stays below the strongest ML results.

Task~1 supports two main observations. The current code snapshot is a strong signal for current-attempt success prediction. Revision-aware inputs can help in specific cases, particularly pairwise \textit{XGBoost}, but offer no consistent advantage across model families. The most reliable conclusion is that lexical ML models, especially \textit{XGBoost}, are the most effective and stable in this setting.

\begin{table}[h]
\centering
\caption{Task~1: class-specific and calibration performance}
\setlength{\arrayrulewidth}{0.2mm}
\setlength{\tabcolsep}{3.5pt}
\renewcommand{\arraystretch}{1.1}
\label{tab:task1_class_calibration}
\scriptsize
\begin{tabular}{l||l||c||c||c||c||c||c}
\hline
\multicolumn{2}{c||}{\textbf{Model Configuration}} & \multicolumn{6}{c}{\textbf{Task 1}} \\ \hline
\multicolumn{1}{l|}{\textbf{Regime}} & \textbf{Model} & \multicolumn{1}{l|}{\textbf{$P$}} & \multicolumn{1}{l|}{\textbf{$R$}} & \multicolumn{1}{l|}{\textbf{$TNR$}} & \multicolumn{1}{l|}{\textbf{$BS$}} & \multicolumn{1}{l|}{\textbf{$\mathcal{L}_{\log}$}} & \textbf{$\tau$} \\ \hline \hline
\multicolumn{1}{l|}{\multirow{6}{*}{Cur}} & \textit{LinearSVM} & \multicolumn{1}{l|}{\textbf{0.8750}} & \multicolumn{1}{l|}{0.6364} & \multicolumn{1}{l|}{0.8889} & \multicolumn{1}{l|}{0.1777} & \multicolumn{1}{l|}{0.5289} & 0.5900 \\ \cline{2-8} 
\multicolumn{1}{l|}{} & \textit{XGBoost} & \multicolumn{1}{l|}{0.8095} & \multicolumn{1}{l|}{0.7727} & \multicolumn{1}{l|}{0.7778} & \multicolumn{1}{l|}{0.1535} & \multicolumn{1}{l|}{0.4867} & 0.5600 \\ \cline{2-8} 
\multicolumn{1}{l|}{} & \textit{BiGRU} & \multicolumn{1}{l|}{0.5714} & \multicolumn{1}{l|}{0.3636} & \multicolumn{1}{l|}{0.6667} & \multicolumn{1}{l|}{0.2776} & \multicolumn{1}{l|}{0.7621} & 0.5900 \\ \cline{2-8} 
\multicolumn{1}{l|}{} & \textit{BiLSTM} & \multicolumn{1}{l|}{0.6875} & \multicolumn{1}{l|}{0.5000} & \multicolumn{1}{l|}{0.7222} & \multicolumn{1}{l|}{0.2639} & \multicolumn{1}{l|}{0.7282} & 0.4900 \\ \cline{2-8} 
\multicolumn{1}{l|}{} & \textit{GraphCodeBERT} & \multicolumn{1}{l|}{0.7143} & \multicolumn{1}{l|}{0.4545} & \multicolumn{1}{l|}{0.7778} & \multicolumn{1}{l|}{0.2481} & \multicolumn{1}{l|}{0.7143} & 0.7000 \\ \cline{2-8} 
\multicolumn{1}{l|}{} & \textit{CodeT5+} & \multicolumn{1}{l|}{0.6538} & \multicolumn{1}{l|}{0.7727} & \multicolumn{1}{l|}{0.5000} & \multicolumn{1}{l|}{0.2919} & \multicolumn{1}{l|}{0.9074} & 0.4000 \\ \hline \hline
\multicolumn{1}{l|}{\multirow{6}{*}{Pair}} & \textit{LinearSVM} & \multicolumn{1}{l|}{0.5455} & \multicolumn{1}{l|}{\textbf{1.0000}} & \multicolumn{1}{l|}{0.5000} & \multicolumn{1}{l|}{0.2138} & \multicolumn{1}{l|}{0.6072} & 0.2100 \\ \cline{2-8} 
\multicolumn{1}{l|}{} & \textit{XGBoost} & \multicolumn{1}{l|}{0.7500} & \multicolumn{1}{l|}{\textbf{1.0000}} & \multicolumn{1}{l|}{0.8000} & \multicolumn{1}{l|}{0.1534} & \multicolumn{1}{l|}{0.4289} & 0.1900 \\ \cline{2-8} 
\multicolumn{1}{l|}{} & \textit{BiGRU} & \multicolumn{1}{l|}{0.3333} & \multicolumn{1}{l|}{0.3333} & \multicolumn{1}{l|}{0.6000} & \multicolumn{1}{l|}{0.2439} & \multicolumn{1}{l|}{0.6807} & 0.4900 \\ \cline{2-8} 
\multicolumn{1}{l|}{} & \textit{BiLSTM} & \multicolumn{1}{l|}{0.3333} & \multicolumn{1}{l|}{0.1667} & \multicolumn{1}{l|}{0.8000} & \multicolumn{1}{l|}{0.3190} & \multicolumn{1}{l|}{0.8938} & 0.7600 \\ \cline{2-8} 
\multicolumn{1}{l|}{} & \textit{GraphCodeBERT} & \multicolumn{1}{l|}{0.5000} & \multicolumn{1}{l|}{\textbf{1.0000}} & \multicolumn{1}{l|}{0.4000} & \multicolumn{1}{l|}{0.2402} & \multicolumn{1}{l|}{0.6652} & 0.1800 \\ \cline{2-8} 
\multicolumn{1}{l|}{} & \textit{CodeT5+} & \multicolumn{1}{l|}{0.0000} & \multicolumn{1}{l|}{0.0000} & \multicolumn{1}{l|}{0.8000} & \multicolumn{1}{l|}{0.2699} & \multicolumn{1}{l|}{0.7333} & 0.5800 \\ \hline \hline
\multicolumn{1}{l|}{\multirow{6}{*}{Mul}} & \textit{LinearSVM} & \multicolumn{1}{l|}{0.6667} & \multicolumn{1}{l|}{0.6667} & \multicolumn{1}{l|}{0.8333} & \multicolumn{1}{l|}{0.2170} & \multicolumn{1}{l|}{0.6156} & 0.2100 \\ \cline{2-8} 
\multicolumn{1}{l|}{} & \textit{XGBoost} & \multicolumn{1}{l|}{0.4286} & \multicolumn{1}{l|}{\textbf{1.0000}} & \multicolumn{1}{l|}{0.3333} & \multicolumn{1}{l|}{\textbf{0.1477}} & \multicolumn{1}{l|}{\textbf{0.4199}} & 0.0300 \\ \cline{2-8} 
\multicolumn{1}{l|}{} & \textit{BiGRU} & \multicolumn{1}{l|}{0.2500} & \multicolumn{1}{l|}{0.3333} & \multicolumn{1}{l|}{0.5000} & \multicolumn{1}{l|}{0.2724} & \multicolumn{1}{l|}{0.7381} & 0.5400 \\ \cline{2-8} 
\multicolumn{1}{l|}{} & \textit{BiLSTM} & \multicolumn{1}{l|}{0.5000} & \multicolumn{1}{l|}{0.3333} & \multicolumn{1}{l|}{0.8333} & \multicolumn{1}{l|}{0.2320} & \multicolumn{1}{l|}{0.6570} & 0.4400 \\ \cline{2-8} 
\multicolumn{1}{l|}{} & \textit{GraphCodeBERT} & \multicolumn{1}{l|}{0.0000} & \multicolumn{1}{l|}{0.0000} & \multicolumn{1}{l|}{\textbf{1.0000}} & \multicolumn{1}{l|}{0.2358} & \multicolumn{1}{l|}{0.6723} & 0.3500 \\ \cline{2-8} 
\multicolumn{1}{l|}{} & \textit{CodeT5+} & \multicolumn{1}{l|}{0.5000} & \multicolumn{1}{l|}{0.6667} & \multicolumn{1}{l|}{0.6667} & \multicolumn{1}{l|}{0.2342} & \multicolumn{1}{l|}{0.6615} & 0.4700 \\ \hline \hline
\end{tabular}
\end{table}

\subsection{Task 2: Next-Attempt Success Prediction}
\label{subsec:task2_results}
Task~2 predicts whether the next attempt will be accepted. \tabboxref{tab:task2_ranking}, \tabboxref{tab:task2_decision}, and 
\tabboxref{tab:task2_class_calibration} report ranking, thresholded decision, and class-specific metrics, respectively. The current-only results show a modest but meaningful signal for immediate future success. \textit{LinearSVM} gives the strongest thresholded current-only performance: MCC of 0.4229, $F_1$ of 0.6667, and $BA$ of 0.7167. In terms of ranking, \textit{CodeT5+} and \textit{BiGRU} lead current-only AP/PR-AUC at 0.6045 and 0.6002, respectively. The signal is weaker than in Task~1, since the target depends on the learner's next revision rather than the current verdict.

\begin{table}[h]
\centering
\caption{Task~2: ranking-oriented performance}
\setlength{\arrayrulewidth}{0.2mm}
\setlength{\tabcolsep}{12.7pt}
\renewcommand{\arraystretch}{1.1}
\label{tab:task2_ranking}
\scriptsize
\begin{tabular}{l||l||c||c}
\hline
\multicolumn{2}{c||}{\textbf{Model Configuration}} & \multicolumn{2}{c}{\textbf{Task 2}} \\ \hline
\multicolumn{1}{l|}{\textbf{Regime}} & \textbf{Model} & \multicolumn{1}{c|}{\textbf{AP/PR-AUC}} & \multicolumn{1}{c}{\textbf{ROC-AUC}} \\ \hline \hline
\multicolumn{1}{l|}{\multirow{6}{*}{Cur}} & \textit{LinearSVM} & \multicolumn{1}{l|}{0.5494} & 0.7333 \\ \cline{2-4} 
\multicolumn{1}{l|}{} & \textit{XGBoost} & \multicolumn{1}{l|}{0.5460} & 0.7167 \\ \cline{2-4} 
\multicolumn{1}{l|}{} & \textit{BiGRU} & \multicolumn{1}{l|}{0.6002} & 0.7083 \\ \cline{2-4} 
\multicolumn{1}{l|}{} & \textit{BiLSTM} & \multicolumn{1}{l|}{0.5002} & 0.6750 \\ \cline{2-4} 
\multicolumn{1}{l|}{} & \textit{GraphCodeBERT} & \multicolumn{1}{l|}{0.4027} & 0.5000 \\ \cline{2-4} 
\multicolumn{1}{l|}{} & \textit{CodeT5+} & \multicolumn{1}{l|}{0.6045} & 0.6500 \\ \hline \hline
\multicolumn{1}{l|}{\multirow{6}{*}{Pair}} & \textit{LinearSVM} & \multicolumn{1}{l|}{0.6389} & 0.8333 \\ \cline{2-4} 
\multicolumn{1}{l|}{} & \textit{XGBoost} & \multicolumn{1}{l|}{0.6389} & 0.8333 \\ \cline{2-4} 
\multicolumn{1}{l|}{} & \textit{BiGRU} & \multicolumn{1}{l|}{0.5000} & 0.5556 \\ \cline{2-4} 
\multicolumn{1}{l|}{} & \textit{BiLSTM} & \multicolumn{1}{l|}{0.6667} & 0.6111 \\ \cline{2-4} 
\multicolumn{1}{l|}{} & \textit{GraphCodeBERT} & \multicolumn{1}{l|}{0.6111} & 0.5556 \\ \cline{2-4} 
\multicolumn{1}{l|}{} & \textit{CodeT5+} & \multicolumn{1}{l|}{0.5333} & 0.7222 \\ \hline \hline
\multicolumn{1}{l|}{\multirow{6}{*}{Mul}} & \textit{LinearSVM} & \multicolumn{1}{l|}{\textbf{1.0000}} & \textbf{1.0000} \\ \cline{2-4} 
\multicolumn{1}{l|}{} & \textit{XGBoost} & \multicolumn{1}{l|}{\textbf{1.0000}} & \textbf{1.0000} \\ \cline{2-4} 
\multicolumn{1}{l|}{} & \textit{BiGRU} & \multicolumn{1}{l|}{0.2000} & 0.0000 \\ \cline{2-4} 
\multicolumn{1}{l|}{} & \textit{BiLSTM} & \multicolumn{1}{l|}{0.2000} & 0.0000 \\ \cline{2-4} 
\multicolumn{1}{l|}{} & \textit{GraphCodeBERT} & \multicolumn{1}{l|}{0.3333} & 0.5000 \\ \cline{2-4} 
\multicolumn{1}{l|}{} & \textit{CodeT5+} & \multicolumn{1}{l|}{0.5000} & 0.7500 \\ \hline \hline
\end{tabular}
\end{table}

\begin{table}[h]
\centering
\caption{Task~2: thresholded decision performance}
\setlength{\arrayrulewidth}{0.2mm}
\setlength{\tabcolsep}{11.2pt}
\renewcommand{\arraystretch}{1.1}
\label{tab:task2_decision}
\scriptsize
\begin{tabular}{l||l||c||c||c}
\hline
\multicolumn{2}{c||}{\textbf{Model Configuration}} & \multicolumn{3}{c}{\textbf{Task 2}} \\ \hline
\multicolumn{1}{l|}{\textbf{Regime}} & \textbf{Model} & \multicolumn{1}{l|}{\textbf{MCC}} & \multicolumn{1}{l|}{\textbf{$F_1$}} & \textbf{$BA$} \\ \hline \hline
\multicolumn{1}{l|}{\multirow{6}{*}{Cur}} & \textit{LinearSVM} & \multicolumn{1}{l|}{0.4229} & \multicolumn{1}{l|}{0.6667} & 0.7167 \\ \cline{2-5} 
\multicolumn{1}{l|}{} & \textit{XGBoost} & \multicolumn{1}{l|}{0.3333} & \multicolumn{1}{l|}{0.6250} & 0.6667 \\ \cline{2-5} 
\multicolumn{1}{l|}{} & \textit{BiGRU} & \multicolumn{1}{l|}{0.2000} & \multicolumn{1}{l|}{0.5000} & 0.6000 \\ \cline{2-5} 
\multicolumn{1}{l|}{} & \textit{BiLSTM} & \multicolumn{1}{l|}{0.2582} & \multicolumn{1}{l|}{0.5714} & 0.6333 \\ \cline{2-5} 
\multicolumn{1}{l|}{} & \textit{GraphCodeBERT} & \multicolumn{1}{l|}{0.0000} & \multicolumn{1}{l|}{0.5455} & 0.5000 \\ \cline{2-5} 
\multicolumn{1}{l|}{} & \textit{CodeT5+} & \multicolumn{1}{l|}{0.1491} & \multicolumn{1}{l|}{0.4000} & 0.5667 \\ \hline \hline
\multicolumn{1}{l|}{\multirow{6}{*}{Pair}} & \textit{LinearSVM} & \multicolumn{1}{l|}{0.5000} & \multicolumn{1}{l|}{0.6667} & 0.7500 \\ \cline{2-5} 
\multicolumn{1}{l|}{} & \textit{XGBoost} & \multicolumn{1}{l|}{0.6325} & \multicolumn{1}{l|}{0.7500} & 0.8333 \\ \cline{2-5} 
\multicolumn{1}{l|}{} & \textit{BiGRU} & \multicolumn{1}{l|}{-0.1890} & \multicolumn{1}{l|}{0.4000} & 0.4167 \\ \cline{2-5} 
\multicolumn{1}{l|}{} & \textit{BiLSTM} & \multicolumn{1}{l|}{0.0000} & \multicolumn{1}{l|}{0.0000} & 0.5000 \\ \cline{2-5} 
\multicolumn{1}{l|}{} & \textit{GraphCodeBERT} & \multicolumn{1}{l|}{0.0000} & \multicolumn{1}{l|}{0.0000} & 0.5000 \\ \cline{2-5} 
\multicolumn{1}{l|}{} & \textit{CodeT5+} & \multicolumn{1}{l|}{0.3162} & \multicolumn{1}{l|}{0.5714} & 0.6667 \\ \hline \hline
\multicolumn{1}{l|}{\multirow{6}{*}{Mul}} & \textit{LinearSVM} & \multicolumn{1}{l|}{\textbf{1.0000}} & \multicolumn{1}{l|}{\textbf{1.0000}} & \textbf{1.0000} \\ \cline{2-5} 
\multicolumn{1}{l|}{} & \textit{XGBoost} & \multicolumn{1}{l|}{\textbf{1.0000}} & \multicolumn{1}{l|}{\textbf{1.0000}} & \textbf{1.0000} \\ \cline{2-5} 
\multicolumn{1}{l|}{} & \textit{BiGRU} & \multicolumn{1}{l|}{-1.0000} & \multicolumn{1}{l|}{0.0000} & 0.0000 \\ \cline{2-5} 
\multicolumn{1}{l|}{} & \textit{BiLSTM} & \multicolumn{1}{l|}{-1.0000} & \multicolumn{1}{l|}{0.0000} & 0.0000 \\ \cline{2-5} 
\multicolumn{1}{l|}{} & \textit{GraphCodeBERT} & \multicolumn{1}{l|}{0.0000} & \multicolumn{1}{l|}{0.3333} & 0.5000 \\ \cline{2-5} 
\multicolumn{1}{l|}{} & \textit{CodeT5+} & \multicolumn{1}{l|}{-0.2500} & \multicolumn{1}{l|}{0.0000} & 0.3750 \\ \hline \hline
\end{tabular}
\end{table}

For the ML family, the pairwise regime provides the strongest practical Task~2 result. \textit{XGBoost} reaches MCC of 0.6325, $F_1$ of 0.75, and $BA$ of 0.8333; \textit{LinearSVM} stays balanced with $P=R=0.6667$ and $TNR=0.8333$. The multi-step ML scores are numerically perfect for both models, but this regime has only five test instances with one positive case: a single correctly placed prediction is enough to yield AP/PR-AUC, ROC-AUC, MCC, $F_1$, and $BA$ of 1. The multi-step result is therefore a small-subset check rather than evidence of general superiority. Among the more reliable current-only and pairwise settings, \textit{LinearSVM} gives the lowest $\mathcal{L}_{\log}$ and shares the strongest pairwise ROC-AUC with \textit{XGBoost}.

\begin{table}[h]
\centering
\caption{Task~2: class-specific and calibration performance}
\setlength{\arrayrulewidth}{0.2mm}
\setlength{\tabcolsep}{3.5pt}
\renewcommand{\arraystretch}{1.1}
\label{tab:task2_class_calibration}
\scriptsize
\begin{tabular}{l||l||c||c||c||c||c||c}
\hline
\multicolumn{2}{c||}{\textbf{Model Configuration}} & \multicolumn{6}{c}{\textbf{Task 2}} \\ \hline
\multicolumn{1}{l|}{\textbf{Regime}} & \textbf{Model} & \multicolumn{1}{l|}{\textbf{$P$}} & \multicolumn{1}{l|}{\textbf{$R$}} & \multicolumn{1}{l|}{\textbf{$TNR$}} & \multicolumn{1}{l|}{\textbf{$BS$}} & \multicolumn{1}{l|}{\textbf{$\mathcal{L}_{\log}$}} & \textbf{$\tau$} \\ \hline \hline
\multicolumn{1}{l|}{\multirow{6}{*}{Cur}} & \textit{LinearSVM} & \multicolumn{1}{l|}{0.5556} & \multicolumn{1}{l|}{0.8333} & \multicolumn{1}{l|}{0.6000} & \multicolumn{1}{l|}{0.2301} & \multicolumn{1}{l|}{0.6469} & 0.2300 \\ \cline{2-8} 
\multicolumn{1}{l|}{} & \textit{XGBoost} & \multicolumn{1}{l|}{0.5000} & \multicolumn{1}{l|}{0.8333} & \multicolumn{1}{l|}{0.5000} & \multicolumn{1}{l|}{0.2223} & \multicolumn{1}{l|}{0.7092} & 0.1000 \\ \cline{2-8} 
\multicolumn{1}{l|}{} & \textit{BiGRU} & \multicolumn{1}{l|}{0.5000} & \multicolumn{1}{l|}{0.5000} & \multicolumn{1}{l|}{0.7000} & \multicolumn{1}{l|}{0.2553} & \multicolumn{1}{l|}{0.7044} & 0.6200 \\ \cline{2-8} 
\multicolumn{1}{l|}{} & \textit{BiLSTM} & \multicolumn{1}{l|}{0.5000} & \multicolumn{1}{l|}{0.6667} & \multicolumn{1}{l|}{0.6000} & \multicolumn{1}{l|}{0.2751} & \multicolumn{1}{l|}{0.7517} & 0.5700 \\ \cline{2-8} 
\multicolumn{1}{l|}{} & \textit{GraphCodeBERT} & \multicolumn{1}{l|}{0.3750} & \multicolumn{1}{l|}{\textbf{1.0000}} & \multicolumn{1}{l|}{0.0000} & \multicolumn{1}{l|}{0.2441} & \multicolumn{1}{l|}{0.6810} & 0.3100 \\ \cline{2-8} 
\multicolumn{1}{l|}{} & \textit{CodeT5+} & \multicolumn{1}{l|}{0.5000} & \multicolumn{1}{l|}{0.3333} & \multicolumn{1}{l|}{0.8000} & \multicolumn{1}{l|}{0.2338} & \multicolumn{1}{l|}{0.6604} & 0.4300 \\ \hline \hline
\multicolumn{1}{l|}{\multirow{6}{*}{Pair}} & \textit{LinearSVM} & \multicolumn{1}{l|}{0.6667} & \multicolumn{1}{l|}{0.6667} & \multicolumn{1}{l|}{0.8333} & \multicolumn{1}{l|}{0.2070} & \multicolumn{1}{l|}{0.5898} & 0.2100 \\ \cline{2-8} 
\multicolumn{1}{l|}{} & \textit{XGBoost} & \multicolumn{1}{l|}{0.6000} & \multicolumn{1}{l|}{\textbf{1.0000}} & \multicolumn{1}{l|}{0.6667} & \multicolumn{1}{l|}{0.2260} & \multicolumn{1}{l|}{0.6362} & 0.0400 \\ \cline{2-8} 
\multicolumn{1}{l|}{} & \textit{BiGRU} & \multicolumn{1}{l|}{0.2857} & \multicolumn{1}{l|}{0.6667} & \multicolumn{1}{l|}{0.1667} & \multicolumn{1}{l|}{0.2583} & \multicolumn{1}{l|}{0.7099} & 0.4700 \\ \cline{2-8} 
\multicolumn{1}{l|}{} & \textit{BiLSTM} & \multicolumn{1}{l|}{0.0000} & \multicolumn{1}{l|}{0.0000} & \multicolumn{1}{l|}{\textbf{1.0000}} & \multicolumn{1}{l|}{0.2536} & \multicolumn{1}{l|}{0.7003} & 0.5500 \\ \cline{2-8} 
\multicolumn{1}{l|}{} & \textit{GraphCodeBERT} & \multicolumn{1}{l|}{0.0000} & \multicolumn{1}{l|}{0.0000} & \multicolumn{1}{l|}{\textbf{1.0000}} & \multicolumn{1}{l|}{0.2312} & \multicolumn{1}{l|}{0.6607} & 0.3300 \\ \cline{2-8} 
\multicolumn{1}{l|}{} & \textit{CodeT5+} & \multicolumn{1}{l|}{0.5000} & \multicolumn{1}{l|}{0.6667} & \multicolumn{1}{l|}{0.6667} & \multicolumn{1}{l|}{0.2344} & \multicolumn{1}{l|}{0.6618} & 0.4700 \\ \hline \hline
\multicolumn{1}{l|}{\multirow{6}{*}{Mul}} & \textit{LinearSVM} & \multicolumn{1}{l|}{\textbf{1.0000}} & \multicolumn{1}{l|}{\textbf{1.0000}} & \multicolumn{1}{l|}{\textbf{1.0000}} & \multicolumn{1}{l|}{\textbf{0.1171}} & \multicolumn{1}{l|}{\textbf{0.3737}} & 0.2000 \\ \cline{2-8} 
\multicolumn{1}{l|}{} & \textit{XGBoost} & \multicolumn{1}{l|}{\textbf{1.0000}} & \multicolumn{1}{l|}{\textbf{1.0000}} & \multicolumn{1}{l|}{\textbf{1.0000}} & \multicolumn{1}{l|}{0.1540} & \multicolumn{1}{l|}{0.4284} & 0.0300 \\ \cline{2-8} 
\multicolumn{1}{l|}{} & \textit{BiGRU} & \multicolumn{1}{l|}{0.0000} & \multicolumn{1}{l|}{0.0000} & \multicolumn{1}{l|}{0.0000} & \multicolumn{1}{l|}{0.2333} & \multicolumn{1}{l|}{0.6602} & 0.4300 \\ \cline{2-8} 
\multicolumn{1}{l|}{} & \textit{BiLSTM} & \multicolumn{1}{l|}{0.0000} & \multicolumn{1}{l|}{0.0000} & \multicolumn{1}{l|}{0.0000} & \multicolumn{1}{l|}{0.2568} & \multicolumn{1}{l|}{0.7068} & 0.4700 \\ \cline{2-8} 
\multicolumn{1}{l|}{} & \textit{GraphCodeBERT} & \multicolumn{1}{l|}{0.2000} & \multicolumn{1}{l|}{\textbf{1.0000}} & \multicolumn{1}{l|}{0.0000} & \multicolumn{1}{l|}{0.2646} & \multicolumn{1}{l|}{0.7225} & 0.0100 \\ \cline{2-8} 
\multicolumn{1}{l|}{} & \textit{CodeT5+} & \multicolumn{1}{l|}{0.0000} & \multicolumn{1}{l|}{0.0000} & \multicolumn{1}{l|}{0.7500} & \multicolumn{1}{l|}{0.2207} & \multicolumn{1}{l|}{0.6342} & 0.4700 \\ \hline \hline
\end{tabular}
\end{table}

The DL results are strongest in the current-only regime. \textit{BiGRU} obtains AP/PR-AUC of 0.6002; \textit{BiLSTM} gives the stronger thresholded result with MCC of 0.2582 and $F_1$ of 0.5714. In the pairwise setting, \textit{BiGRU} preserves some positive detection but with weak $TNR$, while \textit{BiLSTM} becomes overly conservative and predicts no positive cases. In the multi-step setting, both models produce negative MCC and zero $F_1$, reflecting unstable threshold transfer on the very small subset. For Task~2, the recurrent models are most reliable in the current-only setting.

The PTMs produce mixed results for Task~2. In the current-only regime, \textit{CodeT5+} gives the strongest ranking among all current-only models, though thresholded performance is weaker. \textit{GraphCodeBERT} achieves perfect $R$ but zero $TNR$, meaning the selected threshold favors positive predictions too strongly. In the pairwise regime, \textit{CodeT5+} provides the most usable PTM result: MCC of 0.3162 and $F_1$ of 0.5714. Pairwise \textit{GraphCodeBERT} predicts no positive cases, giving perfect $TNR$ but zero $R$. In the multi-step regime, \textit{CodeT5+} preserves some ranking signal (ROC-AUC of 0.75), but the threshold does not transfer well to the test set.

Task~2 is the most difficult setting across all three tasks. The current-only and pairwise regimes produce the most meaningful results, with pairwise \textit{XGBoost} giving the strongest practical outcome. The perfect multi-step ML scores reflect small-subset behavior and should not be generalized. Feature-based ML models remain the most stable; DL and PTMs show useful but threshold-sensitive behavior under data sparsity.

\subsection{Task 3: Near-Future Success Prediction}
\label{subsec:task3_results}
Task~3 predicts whether acceptance occurs within the next three available attempts from an unresolved state. Ranking, thresholded decision, and class-specific results are in \tabboxref{tab:task3_ranking}, \tabboxref{tab:task3_decision}, and \tabboxref{tab:task3_class_calibration}, respectively. Compared with Task~2, Task~3 shows a clearer predictive structure since the short recovery window is a less demanding target than immediate next-attempt success. In the current-only regime, \textit{XGBoost} achieves the strongest ranking result: AP/PR-AUC of 0.9909 and ROC-AUC of 0.9875, along with MCC of 0.6325 and $BA$ of 0.8, while 
\textit{GraphCodeBERT} obtains the highest current-only $F_1$ (0.8).

\begin{table}[h]
\centering
\caption{Task~3: ranking-oriented performance}
\setlength{\arrayrulewidth}{0.2mm}
\setlength{\tabcolsep}{12.7pt}
\renewcommand{\arraystretch}{1.1}
\label{tab:task3_ranking}
\scriptsize
\begin{tabular}{l||l||c||c}
\hline
\multicolumn{2}{c||}{\textbf{Model Configuration}} & \multicolumn{2}{c}{\textbf{Task 3}} \\ \hline
\multicolumn{1}{l|}{\textbf{Regime}} & \textbf{Model} & \multicolumn{1}{c|}{\textbf{AP/PR-AUC}} & \multicolumn{1}{c}{\textbf{ROC-AUC}} \\ \hline \hline
\multicolumn{1}{l|}{\multirow{6}{*}{Cur}} & \textit{LinearSVM} & \multicolumn{1}{l|}{0.9354} & 0.9250 \\ \cline{2-4} 
\multicolumn{1}{l|}{} & \textit{XGBoost} & \multicolumn{1}{l|}{0.9909} & 0.9875 \\ \cline{2-4} 
\multicolumn{1}{l|}{} & \textit{BiGRU} & \multicolumn{1}{l|}{0.8651} & 0.7750 \\ \cline{2-4} 
\multicolumn{1}{l|}{} & \textit{BiLSTM} & \multicolumn{1}{l|}{0.9263} & 0.8875 \\ \cline{2-4} 
\multicolumn{1}{l|}{} & \textit{GraphCodeBERT} & \multicolumn{1}{l|}{0.7127} & 0.7750 \\ \cline{2-4} 
\multicolumn{1}{l|}{} & \textit{CodeT5+} & \multicolumn{1}{l|}{0.7645} & 0.7750 \\ \hline \hline
\multicolumn{1}{l|}{\multirow{6}{*}{Pair}} & \textit{LinearSVM} & \multicolumn{1}{l|}{0.9500} & 0.9583 \\ \cline{2-4} 
\multicolumn{1}{l|}{} & \textit{XGBoost} & \multicolumn{1}{l|}{0.9167} & 0.9167 \\ \cline{2-4} 
\multicolumn{1}{l|}{} & \textit{BiGRU} & \multicolumn{1}{l|}{0.5048} & 0.2917 \\ \cline{2-4} 
\multicolumn{1}{l|}{} & \textit{BiLSTM} & \multicolumn{1}{l|}{0.5750} & 0.5000 \\ \cline{2-4} 
\multicolumn{1}{l|}{} & \textit{GraphCodeBERT} & \multicolumn{1}{l|}{0.7153} & 0.7083 \\ \cline{2-4} 
\multicolumn{1}{l|}{} & \textit{CodeT5+} & \multicolumn{1}{l|}{0.8042} & 0.8750 \\ \hline \hline
\multicolumn{1}{l|}{\multirow{6}{*}{Mul}} & \textit{LinearSVM} & \multicolumn{1}{l|}{\textbf{1.0000}} & \textbf{1.0000} \\ \cline{2-4} 
\multicolumn{1}{l|}{} & \textit{XGBoost} & \multicolumn{1}{l|}{\textbf{1.0000}} & \textbf{1.0000} \\ \cline{2-4} 
\multicolumn{1}{l|}{} & \textit{BiGRU} & \multicolumn{1}{l|}{0.2000} & 0.2000 \\ \cline{2-4} 
\multicolumn{1}{l|}{} & \textit{BiLSTM} & \multicolumn{1}{l|}{0.3333} & 0.6000 \\ \cline{2-4} 
\multicolumn{1}{l|}{} & \textit{GraphCodeBERT} & \multicolumn{1}{l|}{0.2500} & 0.4000 \\ \cline{2-4} 
\multicolumn{1}{l|}{} & \textit{CodeT5+} & \multicolumn{1}{l|}{0.5000} & 0.8000 \\ \hline \hline
\end{tabular}
\end{table}

Within the ML family, the strongest stable evidence comes from the current-only 
and pairwise regimes. Current-only \textit{XGBoost} is the most reliable reference, performing consistently across ranking, thresholded, and probabilistic metrics on the largest Task~3 subset. Pairwise \textit{LinearSVM} also performs strongly: MCC of 0.8018, $F_1$ of 0.8571, and $BA$ of 0.8750, with perfect $P$ and $TNR$. Since the pairwise subset is smaller, this is a strong model-specific outcome rather than a general advantage of revision history. In the multi-step regime, \textit{LinearSVM} achieves perfect scores, but with only six test instances and one positive case, a small number of correctly separated predictions yields 1 across all metrics. Multi-step \textit{XGBoost} obtains the lowest $BS$ and $\mathcal{L}_{\log}$, but its threshold is strongly $R$-oriented, with zero $TNR$ and MCC of 0.

\begin{table}[h]
\centering
\caption{Task~3: thresholded decision performance}
\setlength{\arrayrulewidth}{0.2mm}
\setlength{\tabcolsep}{11.2pt}
\renewcommand{\arraystretch}{1.1}
\label{tab:task3_decision}
\scriptsize
\begin{tabular}{l||l||c||c||c}
\hline
\multicolumn{2}{c||}{\textbf{Model Configuration}} & \multicolumn{3}{c}{\textbf{Task 3}} \\ \hline
\multicolumn{1}{l|}{\textbf{Regime}} & \textbf{Model} & \multicolumn{1}{l|}{\textbf{MCC}} & \multicolumn{1}{l|}{\textbf{$F_1$}} & \textbf{$BA$} \\ \hline \hline
\multicolumn{1}{l|}{\multirow{6}{*}{Cur}} & \textit{LinearSVM} & \multicolumn{1}{l|}{0.3953} & \multicolumn{1}{l|}{0.6250} & 0.6875 \\ \cline{2-5} 
\multicolumn{1}{l|}{} & \textit{XGBoost} & \multicolumn{1}{l|}{0.6325} & \multicolumn{1}{l|}{0.7500} & 0.8000 \\ \cline{2-5} 
\multicolumn{1}{l|}{} & \textit{BiGRU} & \multicolumn{1}{l|}{0.4842} & \multicolumn{1}{l|}{0.7059} & 0.7375 \\ \cline{2-5} 
\multicolumn{1}{l|}{} & \textit{BiLSTM} & \multicolumn{1}{l|}{0.4842} & \multicolumn{1}{l|}{0.7059} & 0.7375 \\ \cline{2-5} 
\multicolumn{1}{l|}{} & \textit{GraphCodeBERT} & \multicolumn{1}{l|}{0.5500} & \multicolumn{1}{l|}{0.8000} & 0.7750 \\ \cline{2-5} 
\multicolumn{1}{l|}{} & \textit{CodeT5+} & \multicolumn{1}{l|}{0.4472} & \multicolumn{1}{l|}{0.7368} & 0.7250 \\ \hline \hline
\multicolumn{1}{l|}{\multirow{6}{*}{Pair}} & \textit{LinearSVM} & \multicolumn{1}{l|}{0.8018} & \multicolumn{1}{l|}{0.8571} & 0.8750 \\ \cline{2-5} 
\multicolumn{1}{l|}{} & \textit{XGBoost} & \multicolumn{1}{l|}{0.0000} & \multicolumn{1}{l|}{0.5714} & 0.5000 \\ \cline{2-5} 
\multicolumn{1}{l|}{} & \textit{BiGRU} & \multicolumn{1}{l|}{-0.6124} & \multicolumn{1}{l|}{0.3333} & 0.2500 \\ \cline{2-5} 
\multicolumn{1}{l|}{} & \textit{BiLSTM} & \multicolumn{1}{l|}{0.0000} & \multicolumn{1}{l|}{0.5714} & 0.5000 \\ \cline{2-5} 
\multicolumn{1}{l|}{} & \textit{GraphCodeBERT} & \multicolumn{1}{l|}{0.4082} & \multicolumn{1}{l|}{0.6667} & 0.7083 \\ \cline{2-5} 
\multicolumn{1}{l|}{} & \textit{CodeT5+} & \multicolumn{1}{l|}{0.3563} & \multicolumn{1}{l|}{0.5714} & 0.6667 \\ \hline \hline
\multicolumn{1}{l|}{\multirow{6}{*}{Mul}} & \textit{LinearSVM} & \multicolumn{1}{l|}{\textbf{1.0000}} & \multicolumn{1}{l|}{\textbf{1.0000}} & \textbf{1.0000} \\ \cline{2-5} 
\multicolumn{1}{l|}{} & \textit{XGBoost} & \multicolumn{1}{l|}{0.0000} & \multicolumn{1}{l|}{0.2857} & 0.5000 \\ \cline{2-5} 
\multicolumn{1}{l|}{} & \textit{BiGRU} & \multicolumn{1}{l|}{0.2000} & \multicolumn{1}{l|}{0.3333} & 0.6000 \\ \cline{2-5} 
\multicolumn{1}{l|}{} & \textit{BiLSTM} & \multicolumn{1}{l|}{0.0000} & \multicolumn{1}{l|}{0.2857} & 0.5000 \\ \cline{2-5} 
\multicolumn{1}{l|}{} & \textit{GraphCodeBERT} & \multicolumn{1}{l|}{0.3162} & \multicolumn{1}{l|}{0.4000} & 0.7000 \\ \cline{2-5} 
\multicolumn{1}{l|}{} & \textit{CodeT5+} & \multicolumn{1}{l|}{0.6325} & \multicolumn{1}{l|}{0.6667} & 0.9000 \\ \hline \hline
\end{tabular}
\end{table}

\begin{table}[h]
\centering
\caption{Task~3: class-specific and calibration performance}
\setlength{\arrayrulewidth}{0.2mm}
\setlength{\tabcolsep}{3.5pt}
\renewcommand{\arraystretch}{1.1}
\label{tab:task3_class_calibration}
\scriptsize
\begin{tabular}{l||l||c||c||c||c||c||c}
\hline
\multicolumn{2}{c||}{\textbf{Model Configuration}} & \multicolumn{6}{c}{\textbf{Task 3}} \\ \hline
\multicolumn{1}{l|}{\textbf{Regime}} & \textbf{Model} & \multicolumn{1}{l|}{\textbf{$P$}} & \multicolumn{1}{l|}{\textbf{$R$}} & \multicolumn{1}{l|}{\textbf{$TNR$}} & \multicolumn{1}{l|}{\textbf{$BS$}} & \multicolumn{1}{l|}{\textbf{$\mathcal{L}_{\log}$}} & \textbf{$\tau$} \\ \hline \hline
\multicolumn{1}{l|}{\multirow{6}{*}{Cur}} & \textit{LinearSVM} & \multicolumn{1}{l|}{0.8333} & \multicolumn{1}{l|}{0.5000} & \multicolumn{1}{l|}{0.8750} & \multicolumn{1}{l|}{0.1557} & \multicolumn{1}{l|}{0.4697} & 0.5700 \\ \cline{2-8} 
\multicolumn{1}{l|}{} & \textit{XGBoost} & \multicolumn{1}{l|}{\textbf{1.0000}} & \multicolumn{1}{l|}{0.6000} & \multicolumn{1}{l|}{\textbf{1.0000}} & \multicolumn{1}{l|}{0.0882} & \multicolumn{1}{l|}{0.2665} & 0.6900 \\ \cline{2-8} 
\multicolumn{1}{l|}{} & \textit{BiGRU} & \multicolumn{1}{l|}{0.8571} & \multicolumn{1}{l|}{0.6000} & \multicolumn{1}{l|}{0.8750} & \multicolumn{1}{l|}{0.2324} & \multicolumn{1}{l|}{0.6577} & 0.5600 \\ \cline{2-8} 
\multicolumn{1}{l|}{} & \textit{BiLSTM} & \multicolumn{1}{l|}{0.8571} & \multicolumn{1}{l|}{0.6000} & \multicolumn{1}{l|}{0.8750} & \multicolumn{1}{l|}{0.2163} & \multicolumn{1}{l|}{0.6241} & 0.5800 \\ \cline{2-8} 
\multicolumn{1}{l|}{} & \textit{GraphCodeBERT} & \multicolumn{1}{l|}{0.8000} & \multicolumn{1}{l|}{0.8000} & \multicolumn{1}{l|}{0.7500} & \multicolumn{1}{l|}{0.1894} & \multicolumn{1}{l|}{0.5707} & 0.5000 \\ \cline{2-8} 
\multicolumn{1}{l|}{} & \textit{CodeT5+} & \multicolumn{1}{l|}{0.7778} & \multicolumn{1}{l|}{0.7000} & \multicolumn{1}{l|}{0.7500} & \multicolumn{1}{l|}{0.2416} & \multicolumn{1}{l|}{0.6763} & 0.4900 \\ \hline \hline
\multicolumn{1}{l|}{\multirow{6}{*}{Pair}} & \textit{LinearSVM} & \multicolumn{1}{l|}{\textbf{1.0000}} & \multicolumn{1}{l|}{0.7500} & \multicolumn{1}{l|}{\textbf{1.0000}} & \multicolumn{1}{l|}{0.1323} & \multicolumn{1}{l|}{0.4233} & 0.2900 \\ \cline{2-8} 
\multicolumn{1}{l|}{} & \textit{XGBoost} & \multicolumn{1}{l|}{0.4000} & \multicolumn{1}{l|}{\textbf{1.0000}} & \multicolumn{1}{l|}{0.0000} & \multicolumn{1}{l|}{0.1208} & \multicolumn{1}{l|}{0.4158} & 0.0400 \\ \cline{2-8} 
\multicolumn{1}{l|}{} & \textit{BiGRU} & \multicolumn{1}{l|}{0.2500} & \multicolumn{1}{l|}{0.5000} & \multicolumn{1}{l|}{0.0000} & \multicolumn{1}{l|}{0.3020} & \multicolumn{1}{l|}{0.7995} & 0.4700 \\ \cline{2-8} 
\multicolumn{1}{l|}{} & \textit{BiLSTM} & \multicolumn{1}{l|}{0.4000} & \multicolumn{1}{l|}{\textbf{1.0000}} & \multicolumn{1}{l|}{0.0000} & \multicolumn{1}{l|}{0.2457} & \multicolumn{1}{l|}{0.6847} & 0.0100 \\ \cline{2-8} 
\multicolumn{1}{l|}{} & \textit{GraphCodeBERT} & \multicolumn{1}{l|}{0.6000} & \multicolumn{1}{l|}{0.7500} & \multicolumn{1}{l|}{0.6667} & \multicolumn{1}{l|}{0.2345} & \multicolumn{1}{l|}{0.6619} & 0.4700 \\ \cline{2-8} 
\multicolumn{1}{l|}{} & \textit{CodeT5+} & \multicolumn{1}{l|}{0.6667} & \multicolumn{1}{l|}{0.5000} & \multicolumn{1}{l|}{0.8333} & \multicolumn{1}{l|}{0.2344} & \multicolumn{1}{l|}{0.6618} & 0.4900 \\ \hline \hline
\multicolumn{1}{l|}{\multirow{6}{*}{Mul}} & \textit{LinearSVM} & \multicolumn{1}{l|}{\textbf{1.0000}} & \multicolumn{1}{l|}{\textbf{1.0000}} & \multicolumn{1}{l|}{\textbf{1.0000}} & \multicolumn{1}{l|}{0.0826} & \multicolumn{1}{l|}{0.3001} & 0.3000 \\ \cline{2-8} 
\multicolumn{1}{l|}{} & \textit{XGBoost} & \multicolumn{1}{l|}{0.1667} & \multicolumn{1}{l|}{\textbf{1.0000}} & \multicolumn{1}{l|}{0.0000} & \multicolumn{1}{l|}{\textbf{0.0147}} & \multicolumn{1}{l|}{\textbf{0.1244}} & 0.0700 \\ \cline{2-8} 
\multicolumn{1}{l|}{} & \textit{BiGRU} & \multicolumn{1}{l|}{0.2000} & \multicolumn{1}{l|}{\textbf{1.0000}} & \multicolumn{1}{l|}{0.2000} & \multicolumn{1}{l|}{0.2885} & \multicolumn{1}{l|}{0.7708} & 0.4900 \\ \cline{2-8} 
\multicolumn{1}{l|}{} & \textit{BiLSTM} & \multicolumn{1}{l|}{0.1667} & \multicolumn{1}{l|}{\textbf{1.0000}} & \multicolumn{1}{l|}{0.0000} & \multicolumn{1}{l|}{0.2478} & \multicolumn{1}{l|}{0.6888} & 0.4200 \\ \cline{2-8} 
\multicolumn{1}{l|}{} & \textit{GraphCodeBERT} & \multicolumn{1}{l|}{0.2500} & \multicolumn{1}{l|}{\textbf{1.0000}} & \multicolumn{1}{l|}{0.4000} & \multicolumn{1}{l|}{0.1512} & \multicolumn{1}{l|}{0.4790} & 0.2100 \\ \cline{2-8} 
\multicolumn{1}{l|}{} & \textit{CodeT5+} & \multicolumn{1}{l|}{0.5000} & \multicolumn{1}{l|}{\textbf{1.0000}} & \multicolumn{1}{l|}{0.8000} & \multicolumn{1}{l|}{0.2285} & \multicolumn{1}{l|}{0.6500} & 0.4900 \\ \hline \hline
\end{tabular}
\end{table}

The DL current-only setting is the most reliable for Task~3. \textit{BiGRU} and \textit{BiLSTM} produce identical thresholded performance: MCC of 0.4842, $F_1$ of 0.7059, and $BA$ of 0.7375. \textit{BiLSTM} leads in ranking and calibration: AP/PR-AUC of 0.9263, ROC-AUC of 0.8875, with lower $BS$ and $\mathcal{L}_{\log}$ than \textit{BiGRU}. Pairwise and multi-step inputs are less stable: pairwise \textit{BiGRU} shows negative MCC and zero $TNR$; pairwise \textit{BiLSTM} becomes strongly positive-oriented; and in the multi-step regime, values such as \textit{BiGRU}'s AP/PR-AUC, ROC-AUC, and $P$ all equal to 0.2 reflect the small imbalanced subset. The current-only setting remains the most reliable DL formulation for Task~3.

The PTMs perform better on Task~3 than on the earlier tasks. In the current-only regime, \textit{GraphCodeBERT} gives the strongest PTM thresholded result: MCC of 0.55, $F_1$ of 0.8, and $BA$ of 0.7750, with balanced $P$ and $R$. \textit{CodeT5+} also gives a solid current-only result: MCC of 0.4472 and $F_1$ of 0.7368. In the pairwise regime, both remain usable: \textit{GraphCodeBERT} gives better thresholded balance; \textit{CodeT5+} gives stronger ROC-AUC. In the multi-step regime, \textit{CodeT5+} reaches MCC of 0.6325 and $BA$ of 0.9, but this is again based on the small multi-step subset. Pretrained code representations are more useful for near-future success than for immediate next-attempt prediction, particularly when the current code state is used directly.

Task~3 provides the strongest evidence that programming trajectories carry useful signal for short-term recovery prediction. The current-only regime is the most stable across all model families; pairwise and multi-step inputs yield some strong but subset-sensitive results. Near-future success is predictable, especially with feature-based ML models and selected PTMs, but the results do not support a broad claim that longer revision history consistently improves prediction.

\subsection{Task 3 Sensitivity Analysis}
\label{subsec:task3_sensitivity_results}
The sensitivity analysis tests whether Task~3 findings hold under stricter future-horizon control. S1 (\(h_t \geq 1\)) remains valid for binary prediction with 76 training, 13 validation, and 10 test instances; S3 (\(h_t \geq 3\)) becomes single-class after filtering and is addressed separately. S1 results show that robustness is family-dependent; \tabboxref{tab:task3_s1_sensitivity} reports the full S1 performance. \textit{XGBoost} remains strong: AP/PR-AUC of 0.8875, ROC-AUC of 0.9167, MCC of 0.8165, $F_1$ of 0.8889, and $BA$ of 0.9167, with a balanced class 
profile ($P=0.8$, $R=1$, $TNR=0.8333$), supporting the main Task~3 ML conclusion and confirming it is not driven by terminal states. \textit{BiLSTM} is less stable: perfect $R$ but zero $TNR$ and MCC of 0, reflecting an all-positive operating pattern that captures all positive cases but fails to discriminate negatives. The PTM representative \textit{GraphCodeBERT} remains usable but weaker: AP/PR-AUC of 0.7292, ROC-AUC of 0.75, MCC of 0.25, and $F_1$ of 0.6, indicating retained signal under stricter filtering but reduced decision quality relative to the main Task~3 result.

\begin{table*}[t]
\centering
\caption{Task~3-S1 sensitivity results using representative models from each model family}
\setlength{\arrayrulewidth}{0.2mm}
\setlength{\tabcolsep}{6.7pt}
\renewcommand{\arraystretch}{1.1}
\label{tab:task3_s1_sensitivity}
\scriptsize
\begin{tabular}{l||l||c|c|c|c|c|c|c|c|c|c|c}
\hline
\textbf{Family} & \textbf{Model} & \textbf{AP/PR-AUC} & \textbf{ROC-AUC} & \textbf{MCC} & \(\boldsymbol{F_1}\) & \textbf{BA} & \(\boldsymbol{P}\) & \(\boldsymbol{R}\) & \textbf{TNR} & \(\boldsymbol{BS}\) & \(\boldsymbol{\mathcal{L}_{\log}}\) & \(\boldsymbol{\tau}\) \\ \hline \hline
ML & \textit{XGBoost} & \textbf{0.8875} & \textbf{0.9167} & \textbf{0.8165} & \textbf{0.8889} & \textbf{0.9167} & \textbf{0.8000} & \textbf{1.0000} & \textbf{0.8333} & \textbf{0.2050} & \textbf{0.5995} & 0.0600 \\ \hline
DL & \textit{BiLSTM} & 0.4750 & 0.4167 & 0.0000 & 0.5714 & 0.5000 & 0.4000 & \textbf{1.0000} & 0.0000 & 0.3298 & 0.8593 & 0.2800 \\ \hline
PTM-based & \textit{GraphCodeBERT} & 0.7292 & 0.7500 & 0.2500 & 0.6000 & 0.6250 & 0.5000 & 0.7500 & 0.5000 & 0.2100 & 0.6098 & 0.3200 \\ \hline \hline
\end{tabular}
\end{table*}

The S3 subset became single-class after filtering, with all instances negative across training, validation, and test. A diagnostic \textit{XGBoost} run confirmed the label collapse: ROC-AUC was undefined and thresholded metrics were non-informative. S3 is therefore reported as a data-availability observation; no additional model experiments were conducted since the constraint comes from the filtered label distribution, not model behavior. Overall, the sensitivity analysis supports the main Task~3 ML conclusion while showing that DL and PTM robustness is more sensitive to future-horizon restrictions.

\subsection{Cross-Task and Model-Family Comparison}
\label{subsec:cross_task_model_family_comparison}
This subsection gives a synthesis across task difficulty, input-regime 
behavior, and model-family behavior. AP/PR-AUC, MCC, and $F_1$ are used 
selectively: AP/PR-AUC for ranking quality, MCC as the common decision-quality 
anchor, and $F_1$ for positive-class detection.

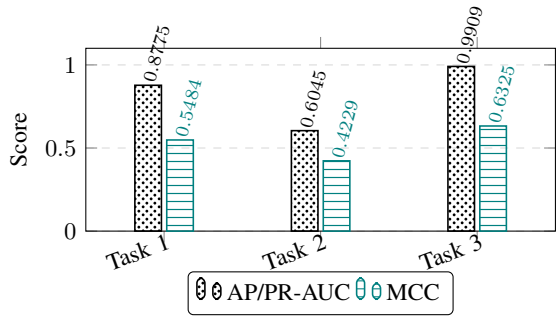
\begin{figure}[h]
\hspace{-3mm}
     \centering
     \captionsetup{justification=centering}
         \begin{tikzpicture}[scale=1]
            \begin{axis}[
    ybar,
    width=0.88\linewidth,
    height=4.0cm,
    bar width=10pt,
    ymin=0,
    ymax=1.10,
    ylabel={Score},
    symbolic x coords={Task 1, Task 2, Task 3},
    xtick=data,
    x tick label style={rotate=20, anchor=east, inner sep=-2pt},
    enlarge x limits=0.25,
    legend style={
        at={(0.5,-0.22)},
        anchor=north,
        legend columns=2,
        draw=black,
        fill=white,
        rounded corners=2pt,
        font=\small
    },
    ymajorgrids=true,
    grid style={dashed,gray!30},
    nodes near coords={\pgfmathprintnumber[fixed,precision=4]{\pgfplotspointmeta}},   
    every node near coord/.append style={font=\scriptsize, rotate=75, anchor=west, inner sep=1pt},
    tick label style={font=\small},
    label style={font=\small},
]
\addplot [color=black, fill=black, semithick, pattern=crosshatch dots, pattern color = black] coordinates {(Task 1,0.8775) (Task 2,0.6045) (Task 3,0.9909)};
\addplot [color=teal, fill=teal!20, semithick, pattern=horizontal lines, pattern color = teal] coordinates {(Task 1,0.5484) (Task 2,0.4229) (Task 3,0.6325)};
\legend{AP/PR-AUC, MCC}
\end{axis}
        \end{tikzpicture}
        \caption{Task difficulty comparison using the strongest current-only AP/PR-AUC and MCC for each task.}
         \label{fig:task_difficulty_comparison}
    \end{figure}

\figboxref{fig:task_difficulty_comparison} summarizes task difficulty using the strongest current-only AP/PR-AUC and MCC for each task. The current-only regime is used because it provides the largest and most stable subset across all tasks. Task~2 is the most difficult setting, with lower best current-only AP/PR-AUC and MCC than Task~1 and Task~3. Task~1 is clearly predictable since the target is tied to the current verdict; Task~3 shows the strongest current-only ranking and decision quality.

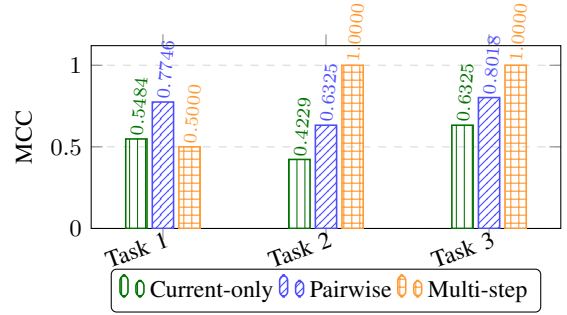
\begin{figure}[h]
\hspace{-3mm}
     \centering
     \captionsetup{justification=centering}
         \begin{tikzpicture}[scale=1]
            \begin{axis}[
    ybar,
    width=0.88\linewidth,
    height=4.0cm,
    bar width=8pt,
    ymin=0,
    ymax=1.12,
    ylabel={MCC},
    symbolic x coords={Task 1, Task 2, Task 3},
    xtick=data,
    x tick label style={rotate=20, anchor=east, inner sep=-2pt},
    enlarge x limits=0.22,
    legend style={
        at={(0.5,-0.22)},
        anchor=north,
        legend columns=3,
        draw=black,
        fill=white,
        rounded corners=2pt,
        font=\small
    },
    ymajorgrids=true,
    grid style={dashed,gray!30},
    point meta=y,
    nodes near coords={\pgfmathprintnumber[fixed,fixed zerofill,precision=4]{\pgfplotspointmeta}},
    every node near coord/.append style={font=\scriptsize, rotate=85, anchor=west, inner sep=1pt},
    tick label style={font=\small},
    label style={font=\small},
]
\addplot [color=green!45!black, fill=green!30, semithick, pattern=vertical lines, pattern color = green!45!black] coordinates {(Task 1,0.5484) (Task 2,0.4229) (Task 3,0.6325)};
\addplot [color=blue!70, fill=blue, semithick, pattern=north east lines, pattern color = blue!70] coordinates {(Task 1,0.7746) (Task 2,0.6325) (Task 3,0.8018)};
\addplot [color=orange!80, fill=orange, semithick, pattern=grid, pattern color = orange!80] coordinates {(Task 1,0.5000) (Task 2,1.0000) (Task 3,1.0000)};
\legend{Current-only, Pairwise, Multi-step}
\end{axis}
        \end{tikzpicture}
        \caption{Input-regime comparison using best MCC per task.}
         \label{fig:regime_behavior_comparison}
    \end{figure}

\figboxref{fig:regime_behavior_comparison} reports the best MCC under current-only, pairwise, and multi-step inputs for each task. Pairwise and multi-step regimes produce several high values, especially in Tasks~1 and~3, but these must be read alongside the smaller eligible subsets in history-based regimes. Multi-step results for Tasks~2 and~3 are particularly sensitive to very small test sets. The comparison does not support a general claim that longer revision history is consistently better; revision-aware inputs provide selected model-specific gains, while current-only remains the most stable formulation across tasks.

\begin{figure}[h]
\hspace{-3mm}
     \centering
     \captionsetup{justification=centering}
         \begin{tikzpicture}[scale=1]
            \begin{axis}[
    ybar,
    width=0.88\linewidth,
    height=4.0cm,
    bar width=8pt,
    ymin=0,
    ymax=1.10,
    ylabel={Score},
    symbolic x coords={ML, DL, PTM},
    xtick=data,
    x tick label style={rotate=15, anchor=east, inner sep=-2pt},
    enlarge x limits=0.22,
    legend style={
        at={(0.5,-0.22)},
        anchor=north,
        legend columns=3,
        draw=black,
        fill=white,
        rounded corners=2pt,
        font=\small
    },
    ymajorgrids=true,
    grid style={dashed,gray!30},
    point meta=y,
    nodes near coords={\pgfmathprintnumber[fixed,fixed zerofill,precision=4]{\pgfplotspointmeta}},
    every node near coord/.append style={font=\scriptsize, rotate=85, anchor=west, inner sep=1pt},
    tick label style={font=\small},
    label style={font=\small},
]
\addplot [color=purple, fill=purple, semithick, pattern=dots, pattern color = purple] coordinates {(ML,0.9909) (DL,0.9263) (PTM,0.7645)};
\addplot [color=cyan!70!black, fill=cyan!70!black, semithick, pattern=north west lines, pattern color = cyan!70!black] coordinates {(ML,0.6325) (DL,0.4842) (PTM,0.5500)};
\addplot [color=brown!70, fill=brown!70, semithick, pattern=crosshatch, pattern color = brown!70] coordinates {(ML,0.7500) (DL,0.7059) (PTM,0.8000)};
\legend{AP/PR-AUC, MCC, $F_1$}
\end{axis}
        \end{tikzpicture}
        \caption{Model-family behavior in the Task~3 current-only setting using AP/PR-AUC, MCC, and $F_1$ within each family.}
         \label{fig:model_family_behavior_comparison}
    \end{figure}
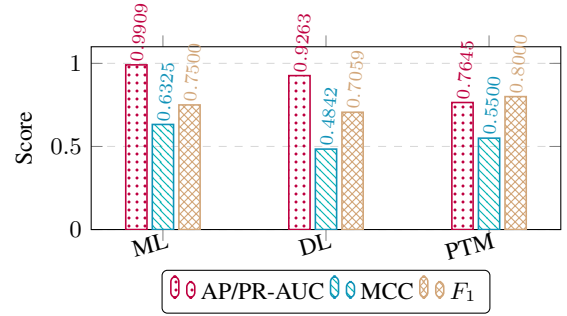

\figboxref{fig:model_family_behavior_comparison} reports the strongest Task~3 
current-only AP/PR-AUC, MCC, and $F_1$ within each family. Task~3 current-only is used because all families show meaningful signal without relying on smaller subsets. ML provides the strongest ranking and decision quality. The PTM family achieves the highest $F_1$, mainly because \textit{GraphCodeBERT} detects positive near-future cases well. The DL family is useful but weaker than ML and PTMs in this setting. ML is the most stable family overall, while DL and PTMs provide narrower but meaningful results, particularly for near-future prediction.

\subsection{Relation to the research questions}
The results provide direct evidence for the four research questions. RQ1 is addressed by the cross-task comparison: programming success is predictable from multi-attempt trajectories, with stronger evidence for current-attempt and near-future success than for next-attempt prediction. RQ2 is addressed by the regime comparison: revision-aware inputs can help selected models but do not consistently improve over current-only representations. RQ3 is addressed by the model-family comparison: ML models are the most stable overall, while DL and PTMs are more task-dependent and most useful in Task~3. Finally, RQ4 is addressed by the Task~3 sensitivity analysis: under stricter future-horizon control, the main Task~3 conclusion remains most robust for the ML family, while DL and PTM representatives are more sensitive to the reduced S1 subset.

\section{Discussion}
\label{sec:discussion}

\subsection{Performance Analysis Across Research Questions}
\label{subsec:discussion_performance_analysis}

\paragraph{RQ1: Predictability of programming success}
Programming success is predictable from multi-attempt trajectories, but prediction strength depends on the horizon. Near-future prediction is informative because an unresolved code state can signal proximity to acceptance, reflected in Task~3's strongest current-only AP/PR-AUC of 0.9909 and MCC of 0.6325 against Task~1's MCC of 0.5484. Task~2 is consistently harder because the next-attempt outcome depends on the learner's subsequent revision decision, debugging strategy, and possible external support, making it less directly tied to the current code state.

\paragraph{RQ2: Effect of revision-aware inputs}
Revision-aware inputs do not provide a consistent advantage over current-only 
representations. Pairwise histories produce strong results in selected cases, such as Task~1 \textit{XGBoost} and Task~3 \textit{LinearSVM}, but these gains are model-dependent and occur on smaller eligible subsets. Multi-step histories are more fragile, with high values tied to very small test splits. The current-only regime is the most stable formulation: it preserves more data and directly represents the code state at the prediction point. Revision history can be useful, but it should not be assumed to improve prediction automatically.

\paragraph{RQ3: Comparison across model families}
ML models provide the most stable behavior across the study, likely because lexical code features are highly informative for judge-system outcomes and can be learned effectively under limited data. \textit{XGBoost} is particularly strong in current-only settings: in Task~3, it reaches AP/PR-AUC of 0.9909 and MCC of 0.6325. The recurrent DL models are more sensitive to data size and input-regime changes, though \textit{BiLSTM} remains meaningful in Task~3 current-only. PTMs provide complementary value, especially for near-future prediction; \textit{GraphCodeBERT} achieves the highest current-only Task~3 $F_1$ of 0.8. However, pretrained code representations do not uniformly outperform ML models under the present dataset size and split structure.

\paragraph{RQ4: Robustness under stricter future-horizon control}
The sensitivity analysis shows that robustness under S1 is family-dependent. \textit{XGBoost} remains strongly discriminative with MCC of 0.8165, while \textit{BiLSTM} loses discrimination entirely (MCC of 0) and 
\textit{GraphCodeBERT} retains partial signal (MCC of 0.25). The S3 subset becomes single-class after filtering, preventing meaningful binary evaluation. Future-horizon availability is therefore not only a data-filtering detail; it directly determines whether near-future prediction can be evaluated fairly.

The results consistently favor simpler, data-efficient modeling choices. Current code states provide the most reliable evidence; ML models remain strong; and revision-aware or pretrained models are most useful only when enough task-specific data is available. The value of trajectory modeling is conditional: revision-aware prediction should be designed with careful attention to horizon definition, sample availability, and model-family stability.

\subsection{Implications for Programming Learning Analytics}
\label{subsec:discussion_programming_learning_analytics}
Judge-system submission traces could support attempt-level learner-state modeling in programming learning analytics. Trajectory data can potentially support three monitoring signals aligned with the study's prediction tasks: a real-time code assessment at submission (Task~1), a next-attempt intervention signal after failure (Task~2), and a recovery alert when acceptance within a short window appears unlikely (Task~3). These signals are particularly valuable in large courses, contests, and OJs where manual inspection of every revision is impractical. The current-only results are especially practical: since the submitted code state provides stable signal without requiring long revision histories, the approach suits settings where trajectories are short or future attempts are unobservable. 

The three tasks correspond to different intervention points. Task~1 provides a real-time code assessment signal at the moment of submission. Task~3 generates a recovery alert that can flag learners unlikely to reach acceptance within three attempts, enabling timely support before repeated failure. Task~2 should be used more cautiously since next-attempt success depends on the learner's subsequent revision strategy and possible external assistance. From a system-design perspective, feature-based ML models serve as strong practical baselines, while DL and PTMs could be added when sufficient data and resources are available. Such systems should complement rather than replace instructor judgment, providing signals for dashboards, tutoring systems, and adaptive support tools.

\subsection{Implications for Revision-Aware SE Analytics}
\label{subsec:discussion_software_engineering_analytics}
Beyond programming education, revision-aware success prediction could inform SE analytics where code evolves through repeated attempts, patches, or debugging cycles. Its value lies in estimating whether an ongoing revision process is close to a successful outcome, rather than evaluating a single code state in isolation. These implications should be interpreted as methodological rather than direct deployment claims. Revision-aware modeling should remain evidence-driven: longer histories may provide useful context but do not automatically improve prediction when they reduce usable data or introduce heterogeneous revision patterns. Compact current-state representations serve as strong baselines for revision analytics, while history-extended and pretrained models are most valuable when they improve robustness rather than only isolated metric values.

\subsection{Opportunities for Human-AI Programming Support}
\label{subsec:discussion_human_ai_programming}
The AI-permitted setting suggests opportunities for human-AI programming support in systems where learners iteratively submit, receive feedback, revise, and resubmit code. Task~3's recovery alert is particularly actionable in this context: when the model signals that acceptance is unlikely within a three-attempt recovery window, the system can trigger targeted support such as a hint, a code review prompt, or an instructor notification, without waiting for further failures. Task~1's code assessment signal supports real-time feedback at the moment of submission, while revision history should be used selectively where it adds reliable context. Together, these signals could contribute to human-AI programming environments that preserve learner agency while offering well-timed, evidence-based support.

\section{Limitations}
\label{sec:threats_validity}
Several limitations apply to this study. First, reconstructed trajectories may be affected by parsing choices, timestamp extraction, verdict normalization, and language-filtering decisions. Second, acceptance verdicts provide a clear success signal but reflect outcome correctness rather than broader learning quality. Third, the study is based on a small AI-permitted contest-style setting, so generalization to other programming contexts requires caution. Fourth, the task construction, trajectory aggregation, threshold selection, and multi-metric reporting reflect one set of design choices; alternative formulations may highlight different aspects of revision behavior. Finally, several pairwise, multi-step, and sensitivity subsets are small, making isolated high metric values unstable; the main interpretation therefore emphasizes consistent cross-task patterns over single best scores from small subsets.

\section{Conclusion}
\label{sec:conclusion}
This study examined revision-aware success prediction from multi-attempt programming trajectories across three prediction horizons: current-attempt acceptance, next-attempt acceptance, and near-future acceptance within a three-attempt recovery window. Tasks were evaluated under current-only, pairwise, and multi-step regimes using ML, DL, and PTM families within a unified trajectory-level protocol. Programming success is predictable, but reliability depends strongly on the task formulation. Current-attempt and near-future prediction produced the clearest signals; next-attempt prediction was the most difficult. In the strongest main Task~3 current-only setting, \textit{XGBoost} reached AP/PR-AUC of 99.09\% and MCC of 0.6325, while \textit{GraphCodeBERT} achieved the highest $F_1$ of 80.00\%. The sensitivity analysis confirmed that Task~3 conclusions were most robust for ML, with 
\textit{XGBoost} reaching MCC of 0.8165 on the stricter S1 subset. Current-only modeling was the most stable formulation across the study. Pairwise and multi-step histories produced gains in selected cases but no consistent advantage, and became more sensitive to reduced sample size.  ML models, particularly \textit{XGBoost}, were the most stable overall; DL and PTMs offered complementary value mainly for near-future prediction. Feature-based ML models remain highly effective for programming success prediction, with task-level outputs such as real-time code assessment (Task~1) and near-future recovery alerts (Task~3) potentially supporting 
submission-aware analytics in OJ platforms and adaptive programming systems.

Future work can extend this approach to larger course-scale and OJ datasets, richer revision features, prompt-aware human-AI programming traces, and broader programming environments. Combining outcome prediction with explanation or feedback generation is a natural next step, enabling predictive signals to support interpretable feedback in programming learning systems.

\section*{Data availability}
The dataset used in this study can be made available upon reasonable request.

\bibliographystyle{ieeetr}

\bibliography{Main}



 




\vfill

\end{document}